\documentclass[12pt,english,round]{article}
\usepackage[T1]{fontenc}
\usepackage[latin9]{inputenc}
\usepackage{geometry}
\usepackage{color}
\usepackage{float}
\usepackage{amsmath}
\usepackage{amssymb}
\usepackage{graphicx}
\usepackage{setspace}
\makeatletter

\providecommand{\tabularnewline}{\\}
\newcommand{\lyxdot}{.}

\makeatother

\usepackage{babel}
\begin{document}
\title{Numerical Solution of Dynamic Portfolio Optimization with Transaction
Costs }
\author{Yongyang Cai\qquad{}Kenneth L. Judd\qquad{}Rong Xu\thanks{Cai (corresponding author), cai.619@osu.edu, The Ohio State University;
Judd, kennethjudd@mac.com, Hoover Institution \& NBER; Xu, rongxu06@gmail.com.
We thank Gerd Infanger, Sunil Kumar, Walter Murray, Michael Saunders,
Karl Schmedders, Benjamin Van Roy, and participants in the SITE Summer
Workshop 2014 at Stanford University for their helpful comments. Cai
acknowledges support from the Hoover Institution at Stanford University.
This research is part of the Blue Waters sustained-petascale computing
project, which is supported by the National Science Foundation (awards
OCI-0725070 and ACI-1238993) and the State of Illinois. Blue Waters
is a joint effort of the University of Illinois at Urbana-Champaign
and its National Center for Supercomputing Applications. This research
was also supported in part by NIH through resources provided by the
Computation Institute and the Biological Sciences Division of the
University of Chicago and Argonne National Laboratory, under grant
1S10OD018495-01. We give special thanks to the\textcolor{black}{{} HTCondor
team of the University of Wisconsin-Madison }for their support. }}
\maketitle
\begin{abstract}
We apply numerical dynamic programming techniques to solve discrete-time
multi-asset dynamic portfolio optimization problems with proportional
transaction costs and shorting/borrowing constraints. Examples include
problems with multiple assets, and many trading periods in a finite
horizon problem. We also solve dynamic stochastic problems, with a
portfolio including one risk-free asset, an option, and its underlying
risky asset, under the existence of transaction costs and constraints.
These examples show that it is now tractable to solve such problems.
\\
 \textit{Keywords}: Numerical dynamic programming, dynamic portfolio
optimization, transaction cost, no-trade region, option hedging, Epstein-Zin
preferences\\
 \textit{JEL Classification: }C61, C63, G11
\end{abstract}
\pagebreak{}

\section{Introduction}

Dynamic management of portfolios is a critical part of any investment
strategy by individuals and firms. Multi-stage portfolio optimization
problems assume that there are $k$ risky assets and/or a risk-less
asset (``bank account'' paying a fixed interest rate) traded during
a time horizon $[0,T]$, and portfolio adjustments are made at $N$
fixed times in the interval $[0,T]$. Trades are made to maximize
the investor's expected utility over terminal wealth ($T$ is the
terminal time) and/or consumption during $[0,T]$.

Standard theory assumes that there is no cost to rebalancing a portfolio,
but transaction costs are not negligible in real markets. Not only
are there transaction fees such as brokerage expenses, but also the
presence of a bid-ask spread creates a transaction cost for a trader.
In the modern U.S. stock market, the bid-ask spread is becoming smaller
and smaller, but varies across stock prices. If stock prices are high,
then the spread could be less than 0.01\%, but if stock prices are
low, then the spread could be more than 0.1\%. Moreover, if risky
assets for trading are not stocks, then they may have high transaction
costs. For example, if an investor wants to exchange two currencies
such as USD and CNY, then transaction costs could be more than 1\%.
Furthermore, if some risky assets such as housing are not very liquid,
then transaction costs could be even higher.

In standard finance theory, the pricing of an option is dependent
on the assumptions that its underlying risky asset can be traded at
any continuous time without transaction costs and there are no constraints
in shorting or borrowing, so that the option can be replicated by
the underlying risky asset and one safe asset. However, while the
transaction costs may be small in one-time trading, the frequency
of rebalancing is theoretically high, so the total transaction cost
will be high. Moreover, investors face constraints in shorting risky
assets or borrowing cash: they have limits on trading amounts, they
have to pay higher costs for shorting risky assets, and they have
to pay higher interest rates for borrowing cash than what they receive
from saving the same amount of cash in a bank account. Therefore,
any examination of real-world dynamic portfolio management needs to
consider these frictions/constraints.

In this paper, we first examine multi-stage portfolio optimization
problems with proportional transaction costs but no options. If the
major transaction cost is the bid-ask spread, then a proportional
transaction cost is the correct case to study. For simplicity, we
assume that neither shorting risky assets nor borrowing cash is allowed
. In fact, if we assume that non-positive wealth has negative infinite
utility (i.e., power utility) and the logarithm of returns of risky
assets have infinite support in both the low and high end, then the
no-shorting and no-borrowing constraints will hold automatically in
discrete-time analysis. We also solve these problems with stochastic
parameters.

However, in many cases, investors would like to reduce or control
the risk and amount of loss in investment strategies. One way is to
add put options into the set of portfolio assets as a hedging strategy.
A put option is a financial contract that gives its holder the right
to sell a certain amount of an underlying asset at a specified price
(strike price) within a specified time (time to maturity, or expiration
time). Investors may also want to include call options in their portfolio.
A call option is a financial contract that gives its holder the right
to buy a certain amount of an underlying asset at a specified price
(strike price) within a specified time (expiration time). In this
paper, we always assume that the options are of European type, that
is, the options can only be exercised at the expiration time (but
they can still be traded before the expiration time).

Multi-stage portfolio optimization problems with transaction costs
have been studied in many papers. The problem with one risky asset
has been well studied; see, e.g., Zabel (1973), Constantinides (1976,
1986), Gennotte and Jung (1994), and Boyle and Lin (1997). The key
insight is that transaction costs create a ``no-trade region'' (NTR);
that is, no trading is done if the current portfolio is inside the
no-trade region, otherwise the investor trades to some point on the
boundary of the no-trade region. Kamin (1975) considers the case with
only two risky assets. Constantinides (1979) and Abrams and Karmarkar
(1980) establish some properties of the NTR for multiple assets, but
only present numerical examples with one safe and one risky asset.
Brown and Smith (2011) evaluate some heuristic strategies and their
bounds based on simulation, but their method cannot give the optimal
portfolios.

In the continuous-time version, there are many papers about the portfolio
optimization problem with transaction costs with one or two risky
assets; see, e.g., Davis and Norman (1990), Duffie and Sun (1990),
Akian et al. (1996), Oksendal and Sulem (2002), Janecek and Shreve
(2004), Liu (2004), Goodman and Ostrov (2010), and Baccarin and Marazzina
(2014, 2016). Muthuraman and Kumar (2006, 2008) give numerical examples
for at most three risky assets. Muthuraman and Zha (2008) provide
a computational scheme that combines simulation with the boundary
update procedure, and present some computational results with $k\geq3$.
However, the presence of simulation implies that the boundary of NTR
can only be roughly approximated. More importantly, it is challenging
for these continuous-time methods to solve problems with constraints
and stochastic parameters.

To the best of our knowledge, when the number of correlated risky
assets is bigger than three and the number of periods is larger than
five, our numerical dynamic programming (DP) method is the first to
explicitly give good numerical solutions with transaction costs and
general utility functions. Moreover, our DP method can be extended
with the recent advances in sparse grid interpolation (see e.g., Judd
et al. (2014) and Brumm and Scheidegger (2017)), to solve even larger
dynamic portfolio problems.

Moreover, this is the first paper to solve the dynamic portfolio optimization
problems with an option and its underlying asset in the portfolio
when transaction costs and borrowing/shorting constraints are present.
For simplicity, when there is an option in the portfolio, we assume
that the portfolio has one riskless asset paying interest, one option,
and its underlying risky asset. It is simple to extend our problem
to cases with multiple risky assets and/or multiple put/call options
with different expiration times and/or strike prices, or some other
derivatives. We assume that the prices of options are given by some
brokerage agencies, which use the pricing formulas in standard finance
theory without consideration of transaction costs and constraints.
However, investors have to pay the transaction costs and face the
constraints in shorting or borrowing. We also study the impact of
Epstein--Zin preferences (Epstein and Zin, 1989) on the optimal allocations.

The paper is organized as follows. Section \ref{sec:Portfolio-Models}
outlines the portfolio models without options and presents their DP
formulation. Section \ref{sec:Numerical-Examples} shows the numerical
results for examples without options. Section \ref{sec:Pricing-Formulas-for}
reviews the pricing formulas for options in standard finance theory.
Section \ref{sec:DP-Models-option} introduces the DP model of the
portfolio optimization problems with options, and also extends it
to a long-horizon portfolio optimization problem with Epstein-Zin
preferences in which the trading horizon is longer than the expiration
time of options. Section \ref{sec:Examples-option} shows the numerical
results of examples with options. Section \ref{sec:Conclusion} concludes.

\section{Portfolio Models without Options\label{sec:Portfolio-Models}}

Assume that there are $k$ risky assets and one risk-less asset available
for investment. The investor can reallocate the portfolio at $N$
periods in $[0,T]$: $0=t_{0}<t_{1}<\cdots<t_{N-1}<t_{N}=T$, which
will incur transaction costs. For simplicity, we assume that $t_{j}$
are equally separated with a length of time $\Delta t=T/N$. Let $\mathbf{R}=(R_{1},\ldots,R_{k})^{\top}$
be the random one-period gross return vector of the risky assets,
and $R_{f}=\exp(r\Delta t)$ be the one-period return of the risk-free
asset, where $r$ is the annual interest rate. In real-life models,
the risk-less gross return $R_{f}$ and the multivariate distribution
of risky asset returns are stochastic and serially-correlated. Assume
that they are dependent on some stochastic parameters such as the
interest rate $r$, the drift $\mathbf{\mu}$, and the volatility
$\mathbf{\sigma}$ of the risky returns. Let all these parameters
be denoted as a vector $\mathbf{\theta}_{t}$ at time $t$. They could
be discrete Markov chains with a given transition probability matrix
from the previous stage to the current stage, or continuously distributed,
conditional on their previous-stage values. For simplicity, in this
paper we assume that $\mathbf{\theta}_{t}$ is a discrete Markov chain,
and then the transition law of $\mathbf{\theta}_{t}$ can be denoted
by 
\begin{equation}
\mathbf{\theta}_{t+\Delta t}=\mathcal{H}_{t}(\mathbf{\theta}_{t},\mathbf{\epsilon}_{t}),\label{eq:theta_law}
\end{equation}
where $\mathbf{\epsilon}_{t}$ are identically and independently distributed
random variables.

The portfolio fraction for asset $i$ at time $t$ (we will always
use $t$ to mean a time $t_{j}$ which is at the beginning of period
$j$ right before reallocation) is denoted $x_{t,i}$, and let $\mathbf{x}_{t}=(x_{t,1},\ldots,x_{t,k})^{\top}$.
Let $W_{t}$ be the total wealth at time $t$ right before reallocation
(it will be changed after reallocation due to the existence of transaction
costs in reallocation), which is the sum of the values of assets.
The difference between the total wealth and the wealth invested in
risky assets is invested in the risk-free asset. Let $\delta_{t,i}W_{t}$
denote the amount of dollars for buying or selling part of the $i$-th
risky asset at time $t$, expressed as a fraction of wealth, where
$\delta_{t,i}>0$ means buying, and $\delta_{t,i}<0$ means selling.
We assume that $f(\delta_{t,i}W_{t})=\tau|\delta_{t,i}W_{t}|$ with
a constant $\tau>0$ is the transaction cost function for buying or
selling part of the $i$-th risky asset using $\delta_{t,i}W_{t}$
dollars. Thus, right after reallocation, the total wealth becomes
$W_{t}-\sum_{i}f(\delta_{t,i}W_{t})$. The absolute operator creates
a kink which should be avoided in optimization problems. Therefore,
for computational purposes, we let $\delta_{t,i}=\delta_{t,i}^{+}-\delta_{t,i}^{-}$
with $\delta_{t,i}^{+},\delta_{t,i}^{-}\geq0$; we know $|\delta_{t,i}|\leq\delta_{t,i}^{+}+\delta_{t,i}^{-}$
but the optimal solution will make either $\delta_{t,i}^{+}$ or $\delta_{t,i}^{-}$
to be zero as the investor will not buy and sell the same risky asset
at the same time. That is, in the optimal solution, $\mathbf{\delta}_{t}^{+}=(\delta_{t,1}^{+},\ldots,\delta_{t,k}^{+})^{\top}$
is the vector of fractions of wealth $W_{t}$ for buying risky assets,
and $\mathbf{\delta}_{t}^{-}=(\delta_{t,1}^{-},\ldots,\delta_{t,k}^{-})^{\top}$
is the vector of fractions of wealth $W_{t}$ for selling risky assets,
so $|\mathbf{\delta}_{t}|=\mathbf{\delta}_{t}^{+}+\mathbf{\delta}_{t}^{-}$,
where the absolute operator is component-wise.

Let $\mathbf{e}$ denote the column vector of all elements equal to
1. The law of transition of the total wealth is 
\begin{equation}
W_{t+\Delta t}=\mathbf{e}^{\top}\mathbf{X}_{t+\Delta t}+R_{f}(1-\mathbf{e}^{\top}\mathbf{x}_{t}-y_{t})W_{t},\label{eq:law_W}
\end{equation}
where $\mathbf{X}_{t+\Delta t}=(X_{t+\Delta t,1},\ldots,X_{t+\Delta t,k})^{\top}$
is the vector of the amount of dollars invested in the risky assets
at time $t+\Delta t$, i.e., 
\begin{equation}
X_{t+\Delta t,i}=R_{i}(x_{t,i}+\delta_{t,i}^{+}-\delta_{t,i}^{-})W_{t},\label{eq:next-stock}
\end{equation}
and 
\begin{equation}
y_{t}=\mathbf{e}^{\top}(\mathbf{\delta}_{t}^{+}-\mathbf{\delta}_{t}^{-}+\tau(\mathbf{\delta}_{t}^{+}+\mathbf{\delta}_{t}^{-}))\label{eq:transaction_y}
\end{equation}
is the fraction of wealth in buying or selling risky assets with $\mathbf{\delta}_{t}=\mathbf{\delta}_{t}^{+}-\mathbf{\delta}_{t}^{-}$
and associated transaction costs. The next-period portfolio fraction
for asset $i$ becomes 
\begin{equation}
x_{t+\Delta t,i}=X_{t+\Delta t,i}/W_{t+\Delta t}\label{eq:next-fraction}
\end{equation}
for $i=1,...,k$. 

Since we do not allow shorting risky assets, we should have the constraint
\begin{equation}
\mathbf{x}_{t}+\mathbf{\delta}_{t}^{+}-\mathbf{\delta}_{t}^{-}\geq0,\label{eq:no-shorting}
\end{equation}
which means that every element of the vector $(\mathbf{x}_{t}+\mathbf{\delta}_{t}^{+}-\mathbf{\delta}_{t}^{-})$
is nonnegative. The no-borrowing constraint requires 
\begin{equation}
1-\mathbf{e}^{\top}\mathbf{x}_{t}\geq y_{t}.\label{eq:no-borrowing}
\end{equation}

The investor's objective is to maximize expected utility at the terminal
time $T$. Thus, the multi-stage portfolio optimization problem can
be expressed as 
\begin{equation}
V_{0}(W_{0},\mathbf{x}_{0},\mathbf{\theta}_{0})=\max_{\mathbf{\delta}_{t}^{+},\mathbf{\delta}_{t}^{-}\geq0}\text{ \ }\mathbb{E}\left\{ U(W_{T})\right\} ,\label{eq:no-consump-model}
\end{equation}
subject to the constraints (\ref{eq:theta_law})-(\ref{eq:no-borrowing}),
where $\mathbb{E}\{\cdot\}$ is the expectation operator, $U$ is
the utility function, and $\mathbf{\delta}_{t}^{+}$ and $\mathbf{\delta}_{t}^{-}$
are element-wise nonnegative.

We may allow assets to finance consumption during the investment period.
Let the total consumption during period $t$ be $c_{t}W_{t}\Delta t$,
where $c_{t}\Delta t$ is the fraction of wealth $W_{t}$, and $c_{t}W_{t}$
is the annual consumption rate in the period. Then the transition
law of total wealth becomes 
\begin{equation}
W_{t+\Delta t}=\mathbf{e}^{\top}\mathbf{X}_{t+\Delta t}+R_{f}(1-\mathbf{e}^{\top}\mathbf{x}_{t}-y_{t})W_{t}-c_{t}W_{t}\Delta t.\label{eq:law_W_consump}
\end{equation}
Let $\beta=\exp(-\rho\Delta t)$ be the one-period discount factor
and a terminal value function $V_{T}$ is given. The dynamic portfolio
optimization problem is to find optimal transaction $(\mathbf{\delta}_{t}^{+},\mathbf{\delta}_{t}^{-})$
and $c_{t}$ such that we have the maximal expected total utility,
i.e., 
\begin{equation}
V_{0}(W_{0},\mathbf{x}_{0},\mathbf{\theta}_{0})=\max_{\mathbf{\delta}_{t}^{+},\mathbf{\delta}_{t}^{-},c_{t}\geq0}\ \mathbb{E}\left\{ \overset{N-1}{\underset{j=0}{\sum}}\beta^{j}U(c_{t_{j}}W_{t_{j}})\Delta t+\beta^{N}V_{T}(W_{T},\mathbf{x}_{T},\mathbf{\theta}_{T})\right\} ,\label{eq:consump_model}
\end{equation}
subject to the constraints (\ref{eq:theta_law}), (\ref{eq:next-stock})-(\ref{eq:no-shorting}),
(\ref{eq:law_W_consump}), and the following new no-borrowing constraint
\begin{equation}
1-\mathbf{e}^{\top}\mathbf{x}_{t}-c_{t}\Delta t\geq y_{t}.\label{eq:no-borrowing-C}
\end{equation}
In the objective of the maximization problem (\ref{eq:consump_model}),
we choose $U(c_{t_{j}}W_{t_{j}})\Delta t$ instead of $U(c_{t_{j}}W_{t_{j}}\Delta t)$
because (i) when $\Delta t$ is very small, as for CRRA (constant
relative risk aversion) utility functions we use later, $U(c_{t_{j}}W_{t_{j}}\Delta t)$
will make no sense as the marginal utility goes to $-\infty$ when
$\Delta t\rightarrow0$, and (ii) $\overset{N-1}{\underset{j=0}{\sum}}\beta^{j}U(c_{t_{j}}W_{t_{j}})\Delta t$
is an approximation of the integrand: 
\[
\int_{0}^{T}e^{-\rho t}U(c_{t}W_{t})dt
\]
which is standard in continuous time models, whereas $\overset{N-1}{\underset{j=0}{\sum}}\beta^{j}U(c_{t_{j}}W_{t_{j}}\Delta t)$
cannot approximate the integrand.

For dynamic portfolio problems with transaction costs, the continuous
state variables are the total wealth $W_{t}$ and allocation fractions
$\mathbf{x}_{t}$ invested in the risky assets. Here $W_{t}$ and
$\mathbf{x}_{t}$ are the values right before reallocation at time
$t$. Thus, the Bellman equation (Bellman, 1957) of the no-consumption
model (\ref{eq:no-consump-model}) is 
\begin{equation}
V_{t}(W_{t},\mathbf{x}_{t},\mathbf{\theta}_{t})=\max_{\mathbf{\delta}_{t}^{+},\mathbf{\delta}_{t}^{-}\geq0}\text{ \ }\mathbb{E}_{t}\left\{ V_{t+\Delta t}(W_{t+\Delta t},\mathbf{x}_{t+\Delta t},\mathbf{\theta}_{t+\Delta t})\right\} ,\label{eq:DP_no_cons}
\end{equation}
subject to the constraints (\ref{eq:law_W})-(\ref{eq:no-borrowing}),
where $\mathbb{E}_{t}\{\cdot\}$ is the expectation operator conditional
on the information at time $t$. The terminal value function is $V_{T}(W,\mathbf{x},\mathbf{\theta})=U(W)$
for some given utility function $U$.

The Bellman equation of the with-consumption model (\ref{eq:consump_model})
is 
\begin{equation}
V_{t}(W_{t},\mathbf{x}_{t},\mathbf{\theta}_{t})=\max_{\mathbf{\delta}_{t}^{+},\mathbf{\delta}_{t}^{-},c_{t}\geq0}\text{ \ }U(c_{t}W_{t})\Delta t+\beta\mathbb{E}_{t}\left\{ V_{t+\Delta t}(W_{t+\Delta t},\mathbf{x}_{t+\Delta t},\mathbf{\theta}_{t+\Delta t})\right\} ,\label{eq:DP_cons}
\end{equation}
subject to the constraints (\ref{eq:law_W_consump}), (\ref{eq:next-stock})-(\ref{eq:no-shorting}),
and (\ref{eq:no-borrowing-C}). For the model (\ref{eq:DP_cons}),
we assume that all risky assets have to be converted into the risk-less
asset at the terminal time so wealth at time $T$ becomes $(1-\tau\mathbf{e}^{\top}\mathbf{x})W_{T}$
(where $W_{T}$ is the wealth right before conversion). We also assume
that afterwards the investor will save this wealth in a bank account
and always consume only the interest from saving, i.e., $r(1-\tau\mathbf{e}^{\top}\mathbf{x})W_{T}$
for all $t\geq T$, so the terminal value function is 
\begin{equation}
V_{T}(W,\mathbf{x},\mathbf{\theta})=\sum_{j=N}^{\infty}\beta^{j-N}U(r(1-\tau\mathbf{e}^{\top}\mathbf{x})W)\Delta t=\frac{U(r(1-\tau\mathbf{e}^{\top}\mathbf{x})W)\Delta t}{1-\beta},\label{eq:term-val-C}
\end{equation}
assuming that the investor lives forever.

\subsection{Portfolio with a CRRA Utility Function but no Consumption}

\label{subsec:crra} In economics and finance, we usually assume a
CRRA utility function, i.e., $U(W)=W^{1-\gamma}/(1-\gamma)$ for some
constant $\gamma>0$ and $\gamma\neq1$, or $U(W)=\log(W)$ for $\gamma=1$.
Thus, for $U(W)=W^{1-\gamma}/(1-\gamma)$, if we assume that $V_{t+\Delta t}(W_{t+\Delta t},\mathbf{x}_{t+\Delta t},\mathbf{\theta}_{t+\Delta t})=W_{t+\Delta t}^{1-\gamma}\cdot G_{t+\Delta t}(\mathbf{x}_{t+\Delta t},\mathbf{\theta}_{t+\Delta t})$,
then the no-consumption model (\ref{eq:DP_no_cons}) can be transformed
as

\begin{equation}
G_{t}(\mathbf{x}_{t},\mathbf{\theta}_{t})=\max_{\mathbf{\delta}_{t}^{+},\mathbf{\delta}_{t}^{-}\geq0}\text{ \ }\mathbb{E}_{t}\left\{ \Pi_{t+\Delta t}^{1-\gamma}\cdot G_{t+\Delta t}(\mathbf{x}_{t+\Delta t},\mathbf{\theta}_{t+\Delta t})\right\} ,\label{Eq:Port_Power_Model}
\end{equation}
subject to 
\begin{eqnarray}
s_{t+\Delta t,i} & \equiv & R_{i}(x_{t,i}+\delta_{t,i}^{+}-\delta_{t,i}^{-}),\label{eq:def_s}\\
\Pi_{t+\Delta t} & \equiv & \mathbf{e}^{\top}\mathbf{s}_{t+\Delta t}+R_{f}(1-\mathbf{e}^{\top}\mathbf{x}_{t}-y_{t}),\label{eq:def_PI}\\
x_{t+\Delta t,i} & \equiv & s_{t+\Delta t,i}/\Pi_{t+\Delta t},\label{eq:def_next_x}
\end{eqnarray}
and the constraints (\ref{eq:transaction_y}) and (\ref{eq:no-shorting})-(\ref{eq:no-borrowing}).
Moreover, $W_{t+\Delta t}=\Pi_{t+\Delta t}W_{t},$ and $V_{t}(W_{t},\mathbf{x}_{t},\mathbf{\theta}_{t})=W_{t}^{1-\gamma}\cdot G_{t}(\mathbf{x}_{t},\mathbf{\theta}_{t})$
by induction, for any time $t=t_{0},t_{1},\ldots,t_{N}$, while $G_{T}(\mathbf{x},\mathbf{\theta})\equiv1/(1-\gamma)$.
Similarly, for $u(W)=\log(W)$, we can also separate $W_{t}$ and
$(\mathbf{x}_{t},\mathbf{\theta}_{t})$ in the value function $V_{t}(W_{t},\mathbf{x}_{t},\mathbf{\theta}_{t})$.

Since $W_{t}$ and $(\mathbf{x}_{t},\mathbf{\theta}_{t})$ are separable
using the CRRA utility function, we could just do a backward recursion
on the functions $G_{t}(\mathbf{x},\mathbf{\theta})$ instead of $V_{t}(W,\mathbf{x},\mathbf{\theta})$,
so the optimal portfolio rules are independent of wealth $W$. This
transformation makes computation much easier, because it not only
saves one state variable $W_{t}$ but also it avoids the exponentially
expanding domains of $W_{t}$ over time $t$.

We know that there is a ``no-trade region'' (NTR), $\Omega_{t}$,
for any $t=t_{0},t_{1},\ldots,t_{N-1}$. When $\mathbf{x}_{t}\in\Omega_{t}$,
the investor will not trade at all, and when $\mathbf{x}_{t}\notin\Omega_{t}$,
the investor will trade to some point on the boundary of $\Omega_{t}$.
That is, $\Omega_{t}$ is defined as 
\[
\Omega_{t}=\{\mathbf{x}_{t}:\ (\mathbf{\delta}_{t}^{+})^{\ast}=(\mathbf{\delta}_{t}^{-})^{\ast}=0\},
\]
where $(\mathbf{\delta}_{t}^{+})^{\ast}\geq0$ are fractions of wealth
for buying risky assets, and $(\mathbf{\delta}_{t}^{-})^{\ast}\geq0$
are fractions of wealth for selling risky assets, for a given $(\mathbf{x}_{t},\mathbf{\theta}_{t})$.
With a CRRA utility function, the NTR $\Omega_{t}$ is independent
of $W_{t}$ because of the separability of $W_{t}$ and $(\mathbf{x}_{t},\mathbf{\theta}_{t})$.

Abrams and Karmarkar (1980) show that the NTR is a connected set and
that it is a cone when the utility function is assumed to be positively
homogeneous (a function $U(x)$ is positively homogeneous if there
exists a positive value function $\psi(x)$ such that $U(ax)=\psi(a)U(x)$
for any $a>0$). Moreover, in the case of proportional transaction
costs and concave utility functions, the NTR can take on many forms
ranging from a simple half-line to a non-convex set. For further discussion,
see Kamin (1975), Constantinides (1976, 1979, 1986), Davis and Norman
(1990), and Muthuraman and Kumar (2006), among other papers.

\subsection{Portfolio with Transaction Costs and Consumption}

We may allow assets to finance consumption during the investment period.
As we discussed in Section \ref{subsec:crra}, if the utility function
is $U(c)=c^{1-\gamma}/(1-\gamma)$ with $\gamma>0$ and $\gamma\neq1$,
and the terminal value function is $V_{T}(W,\mathbf{x},\mathbf{\theta})=W^{1-\gamma}\cdot G_{T}(\mathbf{x},\mathbf{\theta})$
for some given $G_{T}(\mathbf{x},\mathbf{\theta})$, then we have
$V_{t}(W_{t},\mathbf{x}_{t},\mathbf{\theta}_{t})=W_{t}^{1-\gamma}\cdot G_{t}(\mathbf{x}_{t},\mathbf{\theta}_{t}),$
and

\begin{equation}
G_{t}(\mathbf{x}_{t},\mathbf{\theta}_{t})=\max_{c_{t},\mathbf{\delta}_{t}^{+},\mathbf{\delta}_{t}^{-}\geq0}\ U(c_{t})\Delta t+\beta\mathbb{E}_{t}\left\{ \Pi_{t+\Delta t}^{1-\gamma}\cdot G_{t+\Delta t}(\mathbf{x}_{t+\Delta t},\mathbf{\theta}_{t+\Delta t})\right\} ,\label{Eq:Port_Power_Cons_Model}
\end{equation}
subject to 
\[
\Pi_{t+\Delta t}\equiv\mathbf{e}^{\top}\mathbf{s}_{t+\Delta t}+R_{f}(1-\mathbf{e}^{\top}\mathbf{x}_{t}-y_{t}-c_{t}\Delta t),
\]
\[
1-\mathbf{e}^{\top}\mathbf{x}_{t}-c_{t}\Delta t\geq y_{t},
\]
and the constraints (\ref{eq:transaction_y}), (\ref{eq:theta_law}),
(\ref{eq:no-shorting}), (\ref{eq:def_s}), and (\ref{eq:def_next_x}).
From the terminal value function (\ref{eq:term-val-C}), we have  
\begin{equation}
G_{T}(\mathbf{x},\mathbf{\theta}):=\frac{(r(1-\tau\mathbf{e}^{\top}\mathbf{x}))^{1-\gamma}\Delta t}{(1-\gamma)(1-\beta)}.\label{eq:term-val-G}
\end{equation}
Similarly, we can also have the separability of $W$ and $(\mathbf{x},\mathbf{\theta})$
when $U(c)=\log(c)$ and $V_{T}(W,\mathbf{x},\mathbf{\theta})=\log(W)+G_{T}(\mathbf{x},\mathbf{\theta})$.
Here, the no-trade region $\Omega_{t}$ is defined as 
\[
\Omega_{t}=\{\mathbf{x}_{t}/(1-c_{t}^{\ast}\Delta t):\ (\mathbf{\delta}_{t}^{+})^{\ast}=(\mathbf{\delta}_{t}^{-})^{\ast}=0\},
\]
where $c_{t}^{\ast}\Delta t$ is the optimal fraction of wealth for
consumption, $(\mathbf{\delta}_{t}^{+})^{\ast}\geq0$ are optimal
fractions of wealth for buying risky assets, and $(\mathbf{\delta}_{t}^{-})^{\ast}\geq0$
are optimal fractions of wealth for selling risky assets, for a given
$(\mathbf{x}_{t},\mathbf{\theta}_{t})$.

\subsection{Computational Method\label{subsec:Computational-Method}}

Since we do not allow shorting risky assets or borrowing cash, the
domain of $\mathbf{x}_{t}$ is a simplex: $\{\mathbf{x}\in\mathbb{R}_{+}^{k}:\sum_{i=1}^{k}x_{i}\leq1\}$.
However, for computational convenience, we choose the hypercube $[0,1]^{k}$
as the approximation domain of $\mathbf{x}_{t}$. That is, we assume
that at every period right before transactions, negative cash can
exist, but after transactions cash must be nonnegative so some risky
assets must be sold to cover the negative cash.

In our numerical DP method, we choose degree-$d$ complete Chebyshev
polynomials to approximate value functions $G_{t}(\mathbf{x},\mathbf{\theta})$
(we approximate $V_{t}(W_{t},\mathbf{x}_{t},\mathbf{\theta}_{t})$
for problems without CRRA utility; but in this paper all examples
use CRRA utility), and we always use $(d+1)^{k}$ tensor grids of
Chebyshev nodes (i.e., each state dimension has $d+1$ Chebyshev nodes)
as approximation nodes on the hypercube approximation domain $[0,1]^{k}$
for constructing Chebyshev coefficients using the Chebyshev regression
algorithm (see Appendix \ref{sec:Numerical-DP-Algorithms}). The degree
$d$ or the number of quadrature nodes is chosen such that a higher
degree or a higher number of quadrature nodes has little effect on
the numerical solution. In our numerical examples, we use degree-100
complete Chebyshev polynomials for no-consumption problems (\ref{Eq:Port_Power_Model}),
or degree-60 for with-consumption problems (\ref{Eq:Port_Power_Cons_Model}).
The high degree polynomials are required because the value functions
have kinks on the borders of the NTRs, while the utility of consumption
and the discount factor in the with-consumption model (\ref{Eq:Port_Power_Cons_Model})
make the kinks have a less serious impact on the numerical solutions,
thus requiring a lower-degree approximation than in the no-consumption
model (\ref{Eq:Port_Power_Model}).

We apply the multi-dimensional product Gauss-Hermite quadrature rule
for approximating the expectation in (\ref{Eq:Port_Power_Model})
or (\ref{Eq:Port_Power_Cons_Model}). If the time step size $\Delta t$
is one month (i.e., $\Delta t=1/12$ years) or smaller, then we find
that in all of our numerical examples, using 3 Gauss-Hermite quadrature
nodes for each risky return already achieves an accurate estimate
for the expectation in (\ref{Eq:Port_Power_Model}) or (\ref{Eq:Port_Power_Cons_Model}),
and a higher number of Gauss-Hermite quadrature nodes (e.g., 5, 7,
or 9) has almost no change to the NTRs. This happens because with
a small time step, the variance of a one-period risky return is small.
Moreover, when $\tau=0$, the value functions are independent of $\mathbf{x}_{t}$,
so a small $\tau$ will make the value functions not have high curvatures
except at the border of the NTRs. Thus, a small number of quadrature
nodes can achieve high accuracy. When the time step is one year, we
find it requires 5 quadrature nodes in each dimension as the variance
of a one-year risky return is not small.

The maximization problem in (\ref{Eq:Port_Power_Model}) or (\ref{Eq:Port_Power_Cons_Model})
is solved with the NPSOL optimization package (Gill et al., 1994)
for each approximation node. For each approximation node, we solve
the maximization problem in (\ref{Eq:Port_Power_Model}) or (\ref{Eq:Port_Power_Cons_Model}).
The maximization problems are independent of each other among the
approximation nodes, and there are many approximation nodes in our
portfolio problems. Thus we use the parallel DP algorithm (Cai et
al., 2015), which was developed under the\textcolor{black}{{} HTCondor
system of the University of Wisconsin-Madison at first and was adapted
to supercomputers} \textcolor{black}{later}. For small problems, we
use a Mac Pro with a 3.5 GHz 6-Core Intel Xeon E5. For large problems,
we use the Blue Waters supercomputer.

Appendix \ref{sec:Numerical-DP-Algorithms} provides more discussion
on the numerical DP method, including numerical approximation and
numerical integration. More details of numerical DP can also be found
in Cai (2010, 2019), Judd (1998), and Rust (2008).

\subsection{Portfolio without Transaction Costs}

If we can trade the assets at any continuous time without transaction
costs and the time horizon is infinite, and if we assume the power
utility with $\gamma$ denoting the relative risk aversion, the theoretically
optimal portfolio is the Merton point (Merton 1969, 1971): 
\begin{equation}
\left(\Lambda\Sigma\Lambda\right)^{-1}(\mu-r)/\gamma,\label{eq:Merton_point}
\end{equation}
where $\Lambda$ is a diagonal matrix with the diagonal elements $\Lambda_{ii}=\sigma_{i}$
and $\Sigma$ is the correlation matrix of the risky asset returns.
If the problem includes consumption, then the optimal consumption
can be given by a function of the Merton point (see Samuelson (1969)
and Cai (2010)). Thus, we can use this to check the accuracy of our
numerical DP algorithm in solving the multi-asset portfolio problems
with zero transaction costs, by checking if the NTR converges to the
Merton point when the transaction cost $\tau\rightarrow0$, the horizon
is large enough, and the time step size is small enough for approximating
the infinite-horizon continuous-time problem without transaction costs.
This is alternative to the standard checks using a higher degree approximation
(and a higher number of approximation nodes) or a higher number of
quadrature nodes discussed in Section \ref{subsec:Computational-Method}.

\section{Examples without Options\label{sec:Numerical-Examples}}

In this section, we give several numerical examples for solving multi-stage
portfolio optimization problems with proportional transaction costs
and the power utility function $u(W)=W^{1-\gamma}/(1-\gamma)$. In
these examples, the one-period stochastic return vector $\mathbf{R}$
is always assumed to be log-normal with 
\[
\log(\mathbf{R})\sim\mathcal{N}((\mathbf{\mu}-\frac{\mathbf{\sigma}^{2}}{2})\Delta t,(\mathbf{\Lambda\Sigma\Lambda})\Delta t)
\]
in $\mathbb{R}^{k}$, where $\mathbf{\mu}=(\mu_{1},\cdots,\mu_{k})^{\top}$
is the drift, $\mathbf{\sigma}=(\sigma_{1},\cdots,\sigma_{k})^{\top}$
is the volatility, $\mathbf{\Sigma}$ is the correlation matrix of
the log-returns, $\mathbf{\Lambda}=$ diag$(\sigma_{1},\ldots,\sigma_{k})$,
and $\mathbf{\sigma}^{2}$ is the elementwise square of $\mathbf{\sigma}$.
We can express the returns in terms of the Cholesky factorization
$\mathbf{\Sigma}=\mathbf{L}\mathbf{L}^{\top}$, where $\mathbf{L}=(L_{i,j})_{k\times k}$
is a lower triangular matrix, implying 
\[
\log(R_{i})=(\mu_{i}-\frac{\sigma_{i}^{2}}{2})\Delta t+\sigma_{i}\sqrt{\Delta t}\sum_{j=1}^{i}L_{i,j}z_{j},
\]
where $z_{i}$ are independent standard normal random variables, for
$i=1,\ldots,k$. Therefore, for the optimization problems (\ref{Eq:Port_Power_Model})
and (\ref{Eq:Port_Power_Cons_Model}), we apply a product Gauss-Hermite
quadrature for $\mathbf{R}$ to estimate the conditional expectation
of $\Pi_{t+\Delta t}^{1-\gamma}\cdot G_{t+\Delta t}(\mathbf{x}_{t+\Delta t},\mathbf{\theta}_{t+\Delta t})$.
Here, $r$, $\mathbf{\mu}$, and $\mathbf{\sigma}$ (and even $\mathbf{\Sigma}$)
could be stochastic. That is, $r$, $\mathbf{\mu}$, and $\mathbf{\sigma}$
(and even $\mathbf{\Sigma}$) could be elements of $\mathbf{\theta}_{t}$,
and $R_{f}$ and the distribution of $\mathbf{R}$ could be dependent
on $\mathbf{\theta}_{t}$ in period $t$.

Subsection \ref{subsec:Problems-without-Consumption} uses a three-asset
problem without consumption to illustrate the basic properties of
a NTR and to show that our numerical DP method can achieve high accuracy
in finding the NTR even with daily transactions. Subsection \ref{subsec:Portfolio-with-C}
solves several problems with consumption to show that our numerical
DP method can solve problems with three to five assets at a good accuracy,
and solve problems with stochastic parameters and constraints.

\subsection{Example 1: a Three-Asset Problem without Consumption\label{subsec:Problems-without-Consumption}}

We first solve a three-asset portfolio example of the model (\ref{Eq:Port_Power_Model}),
the multi-stage portfolio optimization problem without consumption.
In this example, the assets available for trading include one risk-free
asset with an interest rate $r$ and $k=2$ uncorrelated risky assets
with log-normal returns. We assume that the utility function at the
terminal time is $U(W)=W^{1-\gamma}/(1-\gamma)$, with $\gamma=3$.
We let $r=0.03$, $\mathbf{\mu}=(0.07,0.07)^{\top}$, and $\mathbf{\sigma}=(0.2,0.2)^{\top}$,
so the Merton point is $(1/3,1/3)$; that is, the optimal portfolio
in the infinite-horizon continuous-time problem without transaction
costs is to invest one third of wealth in each risky asset, and the
remaining one third of wealth in the risk-free asset.

\begin{figure}
\begin{tabular}{cc}
\includegraphics[width=0.5\textwidth]{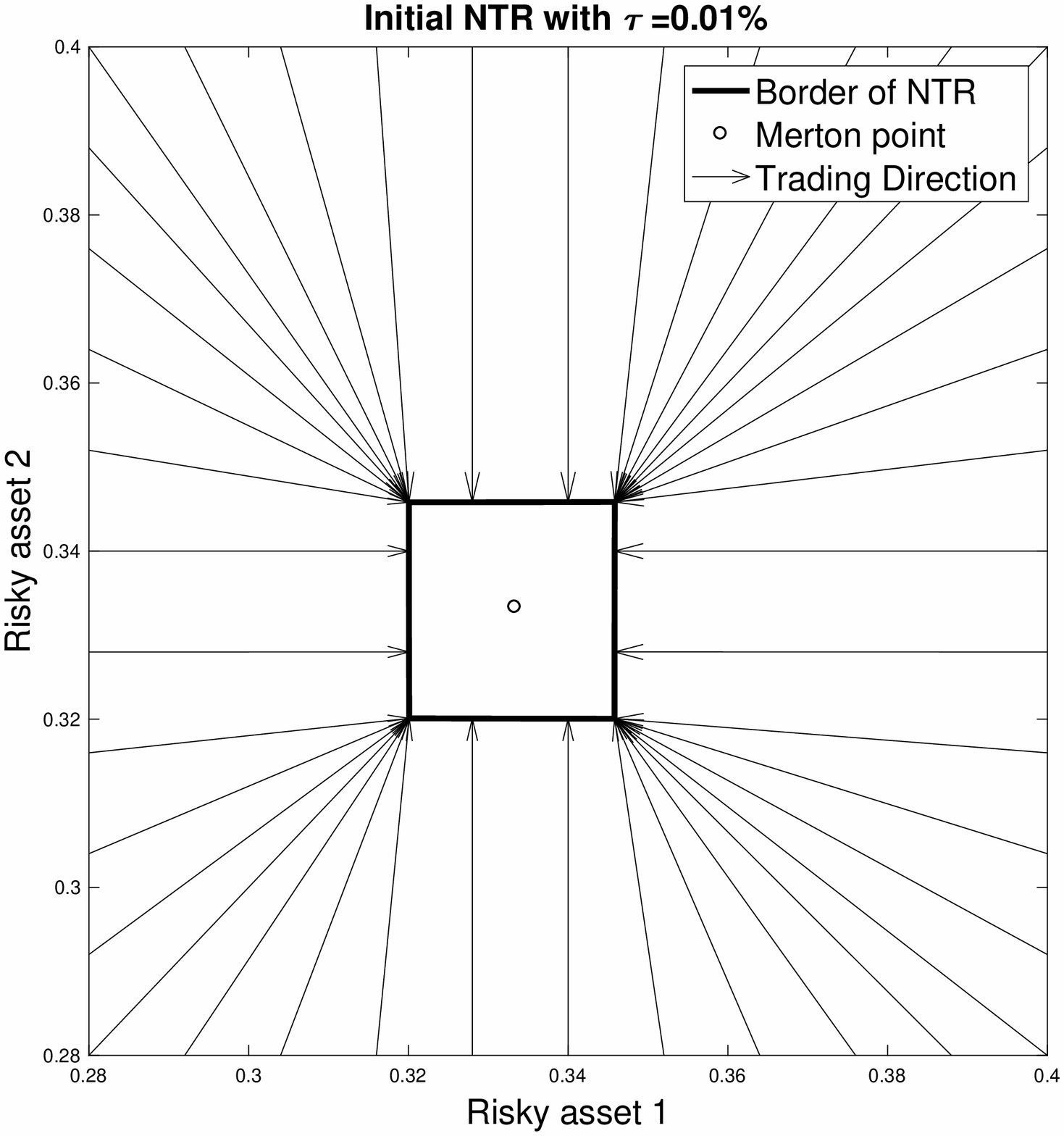}  & \includegraphics[width=0.5\textwidth]{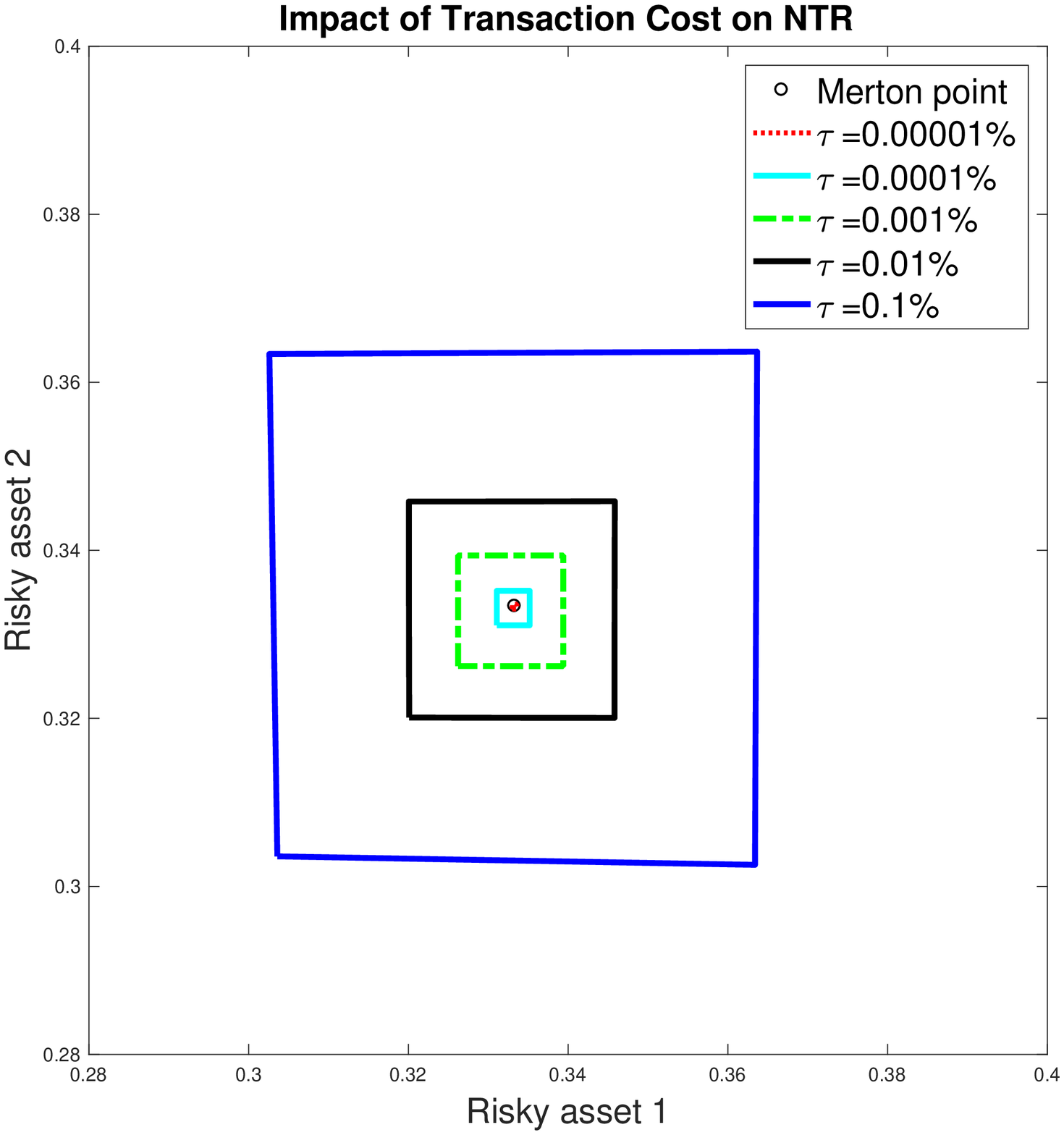}\tabularnewline
\end{tabular}

\caption{NTR at the initial time with various transaction costs. \label{fig:2S_NTR}}
\end{figure}

We let $T=3$ years and choose daily time periods, i.e., $\Delta t=1/365$
years (so the number of periods is $N=1095$). Figure \ref{fig:2S_NTR}
shows the NTRs at the initial time under various transaction costs
ranging from $\tau=0.00001\%$ to $0.1\%$, where horizontal and vertical
axes represent the fraction of wealth invested in risky assets 1 and
2 respectively (all figures in this section use fractions of wealth
invested in risky assets as their axes). The left panel of Figure
\ref{fig:2S_NTR} is for $\tau=0.01\%$. The circle point located
inside the NTR is the Merton point. The NTR is nearly a square and
symmetric w.r.t. the 45 degree line, which is expected because the
two risky assets are i.i.d.. The left panel of Figure \ref{fig:2S_NTR}
also uses the arrows to show the optimal transaction directions. For
example, if the initial portfolio before re-allocation is $(0,0)$,
i.e., all wealth is invested in the risk-less asset, then the optimal
re-allocation is to buy each risky asset with 30.5\% of wealth (the
left-bottom corner of the NTR). Thus, we see that even with the small
transaction cost $\tau=0.01\%$, the NTR is nontrivial: the width
of the NTR is 0.026, i.e., 2.6\% of wealth.

The right panel of Figure \ref{fig:2S_NTR} shows that the NTRs converge
to the Merton point when $\tau\rightarrow0$: when $\tau=0.00001\%$,
the NTR is almost identical to the Merton point. A higher transaction
cost leads to a larger NTR that contains the NTR with smaller transaction
costs. Moreover, as we increase $\tau$, the NTRs grow more slowly
in size than the growth rate of $\tau$, with the width of the NTRs
expanding at a rate of about $\tau^{1/3}$. For example, the width
of the NTR with $\tau=0.1\%$ is 0.061, nearly $2.3$ times the width
with $\tau=0.01\%$, which is close to $\left(0.001/0.0001\right)^{1/3}\approx2.2$.
This is consistent with the theoretical asymptotic result of Goodman
and Ostrov (2010), who found that the width of the NTR is nearly proportional
to $\tau^{1/3}$ for small $\tau$ for infinite-horizon and continuous-time
problems. All these properties of the NTRs verify partially that our
numerical solutions are accurate.

The results in Figure \ref{fig:2S_NTR} are provided by our numerical
DP method, in which we choose degree-100 complete Chebyshev polynomial
approximations, use $101^{2}$ tensor Chebyshev nodes as the approximation
nodes to compute Chebyshev coefficients, and implement the tensor
product Gauss-Hermite quadrature rule with $3^{2}$ tensor quadrature
nodes. We run it on a Mac Pro with a 3.5 GHz 6-Core Intel Xeon E5
using the parallel DP method, and it takes about 1.5 wall clock hours
due to the high degree polynomial approximations. To further verify
that we solve the portfolio problem accurately, we employ degree-150
complete Chebyshev polynomial approximations, $151^{2}$ tensor Chebyshev
nodes, and tensor product Gauss-Hermite quadrature rule with $9^{2}$
tensor quadrature nodes. We find that there is little change in the
solution: e.g., the $\mathcal{L}^{1}$ difference between the degree-100
and the degree-150 solutions is $3.1\times10^{-4}$, and the $\mathcal{L}^{\infty}$
difference is $7.1\times10^{-4}$. A smaller degree polynomial approximation
can be faster but will lead to less accurate solutions. For example,
if we use degree-80 complete Chebyshev polynomial approximations and
$81^{2}$ tensor Chebyshev nodes for the case with $\tau=0.01\%$,
then it takes only 44 minutes, but it also provides a nontrivial change
in the solutions: the $\mathcal{L}^{1}$ difference between the degree-100
and degree-89 solutions is 0.0014, and the $\mathcal{L}^{\infty}$
difference is 0.0033. Note that the kinks on the border of the NTR
in the value and policy functions make it slow to improve the accuracy
of the solution by increasing the degree of polynomial approximation.

\begin{figure}
\begin{tabular}{cc}
\includegraphics[width=0.5\textwidth]{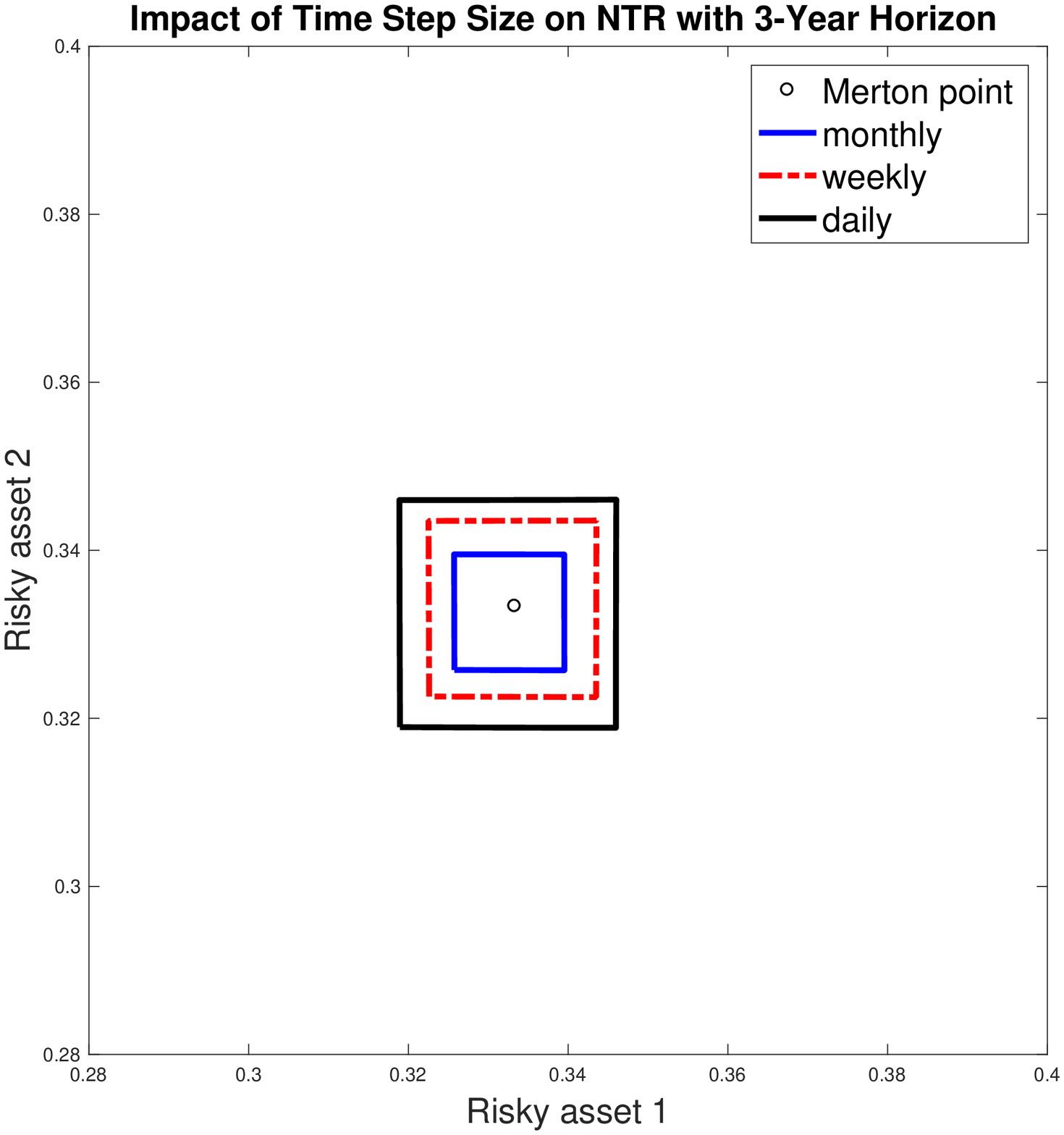} & \includegraphics[width=0.5\textwidth]{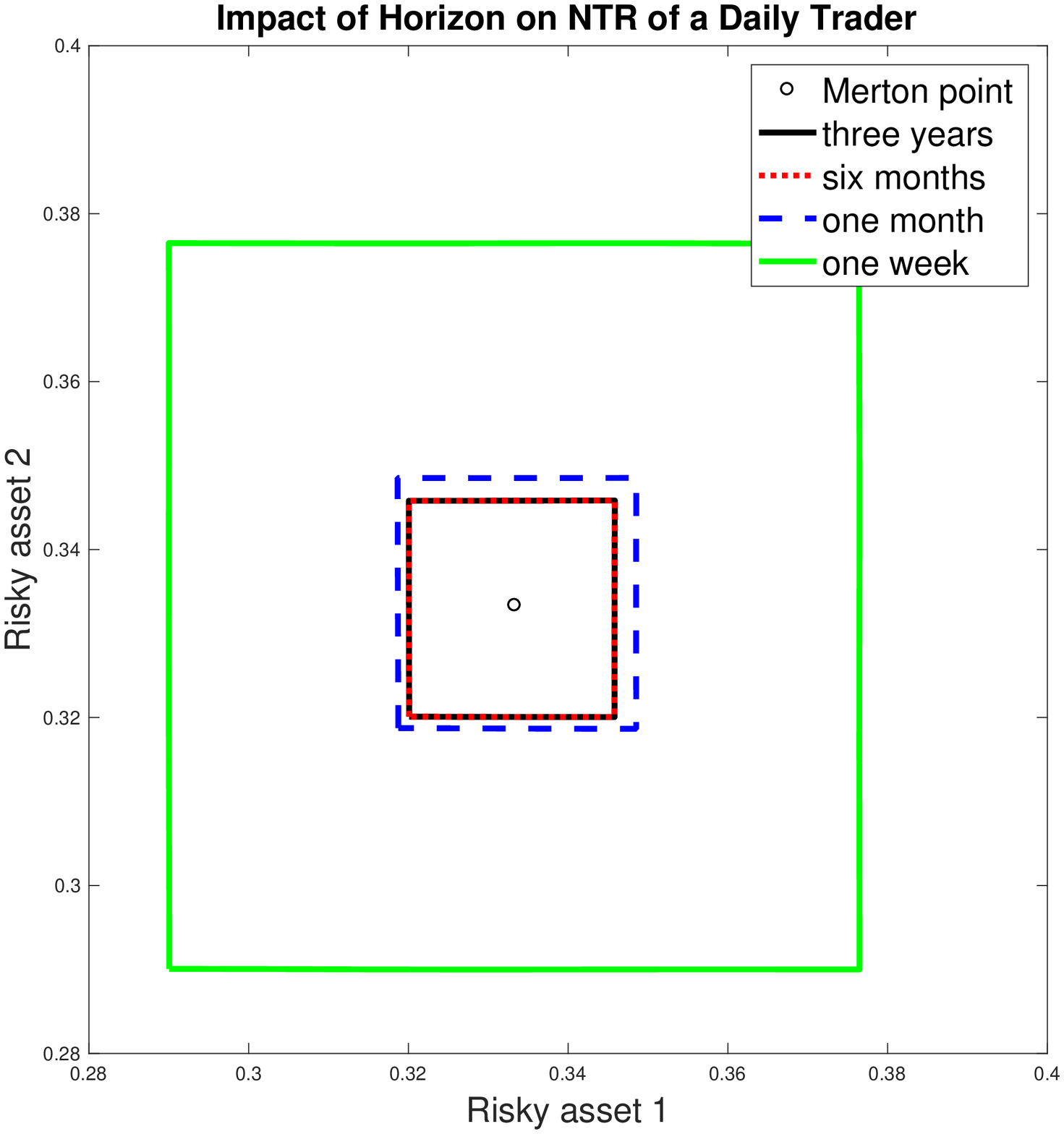}\tabularnewline
\end{tabular}

\caption{Initial NTR with various time step sizes and horizons.\label{fig:2S_T_step}}
\end{figure}

Figure \ref{fig:2S_T_step} shows the NTRs at the initial time with
various time step sizes ranging from daily to monthly (the left panel)
and horizons from one week to three years (the right panel), with
$\tau=0.01\%$. The left panel of Figure \ref{fig:2S_NTR} shows that
the NTR with 3-year horizon expands if the time step size decreases.
The right panel of Figure \ref{fig:2S_T_step} shows that the NTR
of a daily trader shrinks as the horizon increases, and the NTR with
$T=0.6$ years is almost identical to the NTR with $T=3$ years. This
implies that the trading strategy for infinite-horizon continuous-time
problems can be approximated by the initial time solutions of a six-month
horizon problem with daily time steps. This also explains why the
NTRs in the right panel of Figure \ref{fig:2S_NTR} converge to the
Merton point when $\tau\rightarrow0$.

\subsection{Portfolio Problems with Consumption\label{subsec:Portfolio-with-C}}

In this subsection, we solve the with-consumption model (\ref{Eq:Port_Power_Cons_Model}).
The assets available for trading include one risk-free asset with
an interest rate $r$ and multiple risky assets with log-normal annual
returns. We assume that the utility function is $U(c)=c^{1-\gamma}/(1-\gamma)$.
Table \ref{table:parameters_cons} list values of parameters of the
examples in this subsection, where $I_{4}$ represents the $4\times4$
identity matrix. 
\begin{center}
\begin{table}
\begin{centering}
\begin{tabular}{c|c|c|c|c}
\hline 
 & Example 2  & Example 3  & Example 4 & Example 5\tabularnewline
\hline 
$k$  & 2  & 2 & 3  & 4\tabularnewline
$\gamma$  & 2  & 3 & 3  & 4\tabularnewline
$\tau$ & 1\% & 0.1\% & 0.1\% & 0.1\%\tabularnewline
$\beta$  & $\exp(-0.1\Delta t)$  & $\exp(-0.05\Delta t)$ & $\exp(-0.05\Delta t)$  & $\exp(-0.05\Delta t)$ \tabularnewline
$r$  & 0.07  & 0.03 & 0.04  & 0.03\tabularnewline
$\mu^{\top}$  & $(0.15,0.15)$  & stochastic & $(0.07,0.07,0.07)$  & $(0.06,0.066,0.072,0.078)$\tabularnewline
$\sigma^{\top}$  & $(\sqrt{0.17},\sqrt{0.17})$  & $(0.2,0.2)$ & $(0.2,0.2,0.2)$  & $(0.2,0.23,0.26,0.29)$\tabularnewline
$\Sigma$  & $\left[\begin{array}{cc}
1 & 0.4706\\
0.4706 & 1
\end{array}\right]$  & $\left[\begin{array}{cc}
1 & 0\\
0 & 1
\end{array}\right]$ & $\left[\begin{array}{ccc}
1 & 0.4 & 0.4\\
0.4 & 1 & 0.16\\
0.4 & 0.16 & 1
\end{array}\right]$ & $I_{4}$\tabularnewline
\hline 
\end{tabular}
\par\end{centering}
\caption{\label{table:parameters_cons}Parameters for Examples of Portfolio
Problems with Consumption.}
\end{table}
\par\end{center}

\subsubsection{Example 2: a Three-Asset Problem with Consumption}

For the problem with two risky assets and one risk-free asset, the
parameter values are listed in Table \ref{table:parameters_cons},
and they are either the same as or discrete-time analogs to those
in Discussion 1 in Muthuraman and Kumar (2006), including $\tau=1\%$,
in order to show that our numerical DP method can also solve the continuous-time
infinite-horizon portfolio problem with consumption in Muthuraman
and Kumar (2006). We use weekly time steps (i.e., $\Delta t=1/52$
years), and in the numerical DP method for this example, we use degree-$60$
complete Chebyshev polynomials to approximate value functions, use
$61^{2}$ tensor Chebyshev nodes as the approximation nodes to compute
Chebyshev coefficients, and implement the multi-dimensional product
Gauss-Hermite quadrature rule with $3^{2}$ tensor quadrature nodes.
For the case with $T=3$ years (i.e., $N=156$ periods), it takes
only two minutes on a Mac Pro with a 3.5 GHz 6-Core Intel Xeon E5.
A larger degree (e.g., degree-100) polynomial approximation (with
a higher number of Chebyshev nodes) or more quadrature nodes (e.g.,
5 nodes in each dimension) has little impact on the solution: e.g.,
the $\mathcal{L}^{1}$ difference between the degree-60 and degree-100
solutions is $4.6\times10^{-4}$, and the $\mathcal{L}^{\infty}$
difference is 0.0013. In comparison with the no-consumption examples
in Section \ref{subsec:Problems-without-Consumption}, the with-consumption
problems require a much lower degree of polynomial approximation,
because the discount factor and utility of consumption in the objective
of the with-consumption model (\ref{Eq:Port_Power_Cons_Model}) alleviate
the impact of kinks in the value functions on the solutions.

\begin{figure}
\begin{tabular}{cc}
\includegraphics[width=0.5\textwidth]{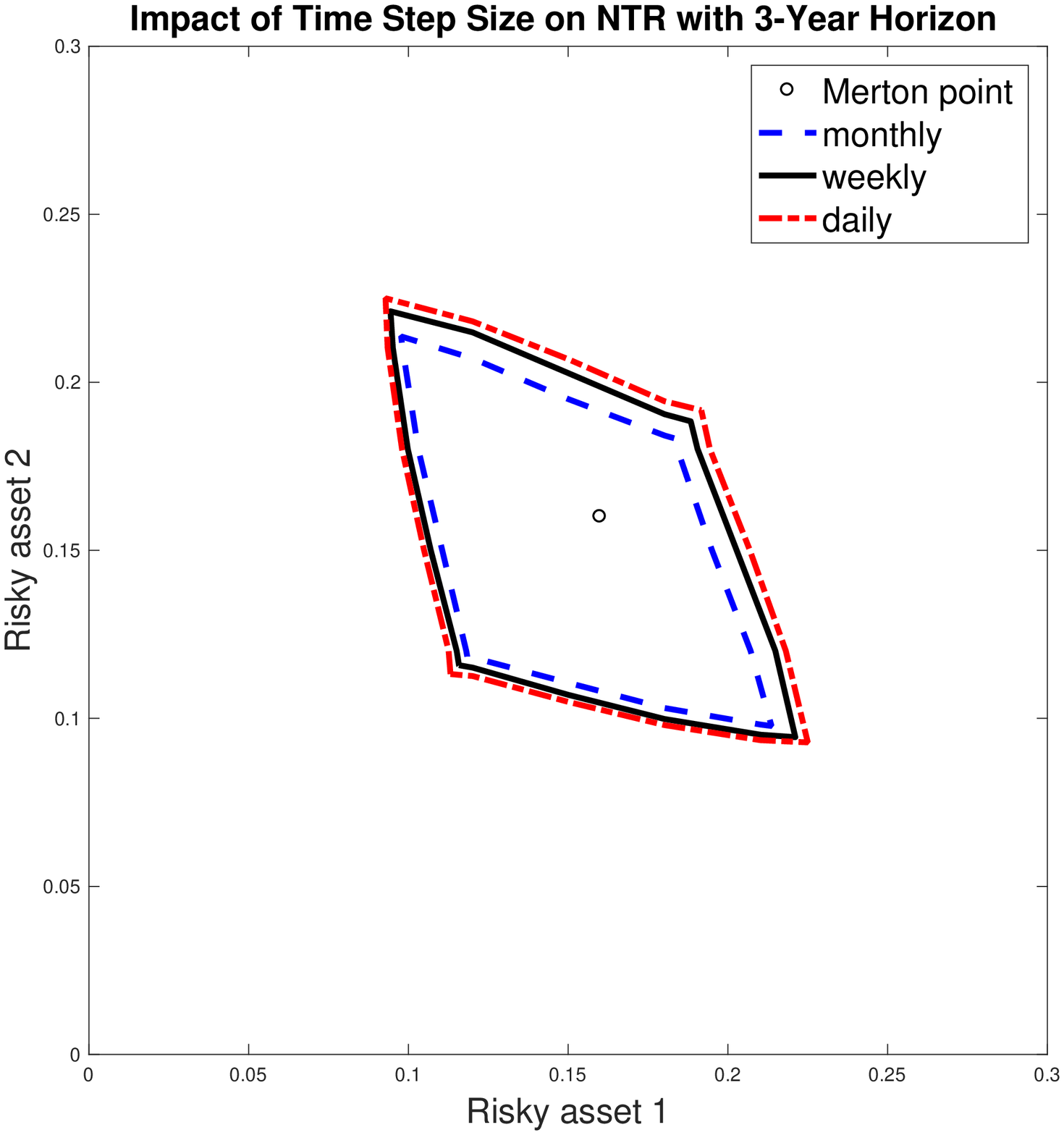}  & \includegraphics[width=0.5\textwidth]{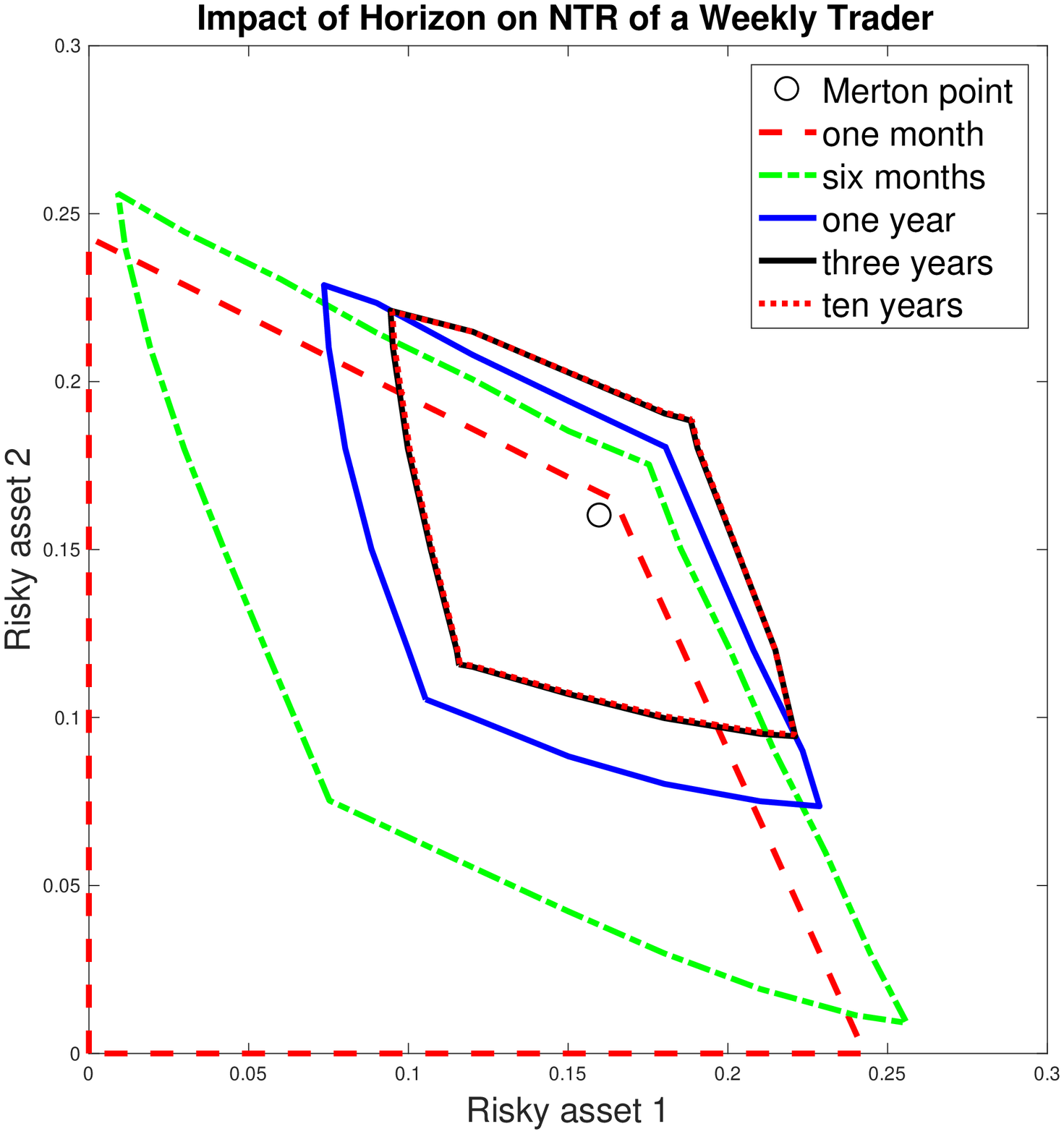}\tabularnewline
\end{tabular}
\centering{}\caption{\label{fig:Kumar_NTR} Impact of time step size and horizon on the
initial-time NTR for 2 correlated risky assets and 1 risk-free asset
with consumption.}
\end{figure}

The left panel of Figure \ref{fig:Kumar_NTR} displays the NTRs at
the initial time with various time step sizes ranging from daily to
monthly for $T=3$ years (i.e., $N=156$ periods). The circle point
located inside the NTRs is the Merton point $(0.16,0.16)$. We see
that the NTR with weekly time steps is close to the NTR with daily
time steps, and is also close to the solution given in Muthuraman
and Kumar (2006) for the infinite-horizon portfolio optimization problems
in which assets can be traded at any continuous time. Thus, we have
shown that our numerical DP can solve continuous-time infinite-horizon
portfolio problem with consumption. Moreover, we can use weekly time
steps to approximate the continuous-time with-consumption model.

The right panel of Figure \ref{fig:Kumar_NTR} shows the initial-time
NTRs with various trading horizons from one month to ten years, for
a fixed time step size of one week. We see that the trading horizon
changes the NTR: a shorter horizon has a larger NTR. We see that the
NTR of the three-year horizon is almost identical to the NTR of the
ten-year horizon. Moreover, when the horizon is small, the no-shorting
constraint becomes binding for some portfolios before rebalancing.
For example, when the horizon has only one month, if the portfolio
before re-allocation is (0,0), then the optimal trading strategy is
to keep the current portfolio unchanged (i.e., the no-shorting constraint
is binding). These occasionally binding constraints make it challenging
for a partial differential equation solution method such as the one
used in Muthuraman and Kumar (2006), but our numerical DP method can
handle them. Clearly, the short-horizon NTRs are impacted by the terminal
condition, i.e., the terminal value function. With our terminal value
function (\ref{eq:term-val-G}) and the assumption to close all risky
assets, NTRs are moving towards the origin if horizon becomes shorter,
as an investor is less willing to keep risky assets so as to avoid
the transaction costs incurred by selling all risky assets at the
terminal time. Thus, the Merton point is no longer in the center of
the NTRs for short-horizon problems and is no longer a good approximate
solution, as its distance to the optimal portfolio may be more than
0.2 for the one-month-horizon problem (as shown in the right panel
of Figure \ref{fig:Kumar_NTR}).

\subsubsection{Example 3: a Three-Asset Problem with Consumption and Stochastic
Drift of Return}

Example 2 has shown that numerical DP can solve a continuous-time
infinite-horizon DP problem, which is often transformed to a Hamilton-Jacobi-Bellman
partial differential equation (PDE) for being solved numerically.
However, if there are stochastic and discrete parameters, then it
is often challenging to solve PDEs with such jump processes. In this
example, we apply numerical DP to solve a three-asset with-consumption
model (\ref{Eq:Port_Power_Cons_Model}), assuming that the drifts
of risky returns, $\mu_{t}=(\mu_{t,1},\mu_{t,2})^{\top}$, are stochastic.
In Appendix \ref{sec:Problems-with-Stochastic}, we also solve dynamic
portfolio problems with stochastic interest rates $r$ or stochastic
volatility $\sigma$. We assume that $\mu_{t,1}$ and $\mu_{t,2}$
are discrete Markov chains and independent of each other. Each $\mu_{t,i}$
has two possible values: $\mu^{1}=0.06$ and $\mu^{2}=0.08$, and
its transition probability matrix is 
\[
\left[\begin{array}{cc}
0.75 & 0.25\\
0.25 & 0.75
\end{array}\right]
\]
for each $i=1,2$, where its $(j_{1},j_{2})$ element represents the
transition probability from $\mu_{t,i}=\mu^{j_{1}}$ to $\mu_{t+\Delta t,i}=\mu^{j_{2}}$,
for $j_{1},j_{2}=1,2$. The other parameter values are listed in Table
\ref{table:parameters_cons}.

We use weekly time periods (i.e., $\Delta t=1/52$ years) in $T=3$
years (so the number of periods is $N=156$). The assets available
for trading include one risk-free asset with an interest rate $r$
and $k=2$ uncorrelated risky assets with log-normal returns. In the
numerical DP method, we choose degree-$60$ complete Chebyshev polynomials
to approximate value functions, use $61^{2}$ tensor Chebyshev nodes
as the approximation nodes, and implement the multi-dimensional product
Gauss-Hermite quadrature rule with $3^{2}$ tensor quadrature nodes.

\begin{figure}
\begin{centering}
\includegraphics[width=0.5\textwidth]{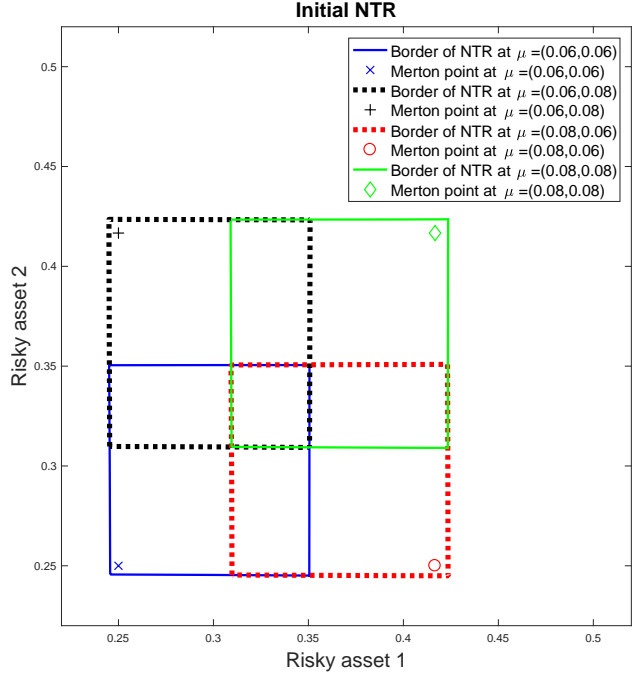} 
\par\end{centering}
\caption{\label{fig:StoPara_mu}NTRs with a stochastic $\mu$.}
\end{figure}

Figure \ref{fig:StoPara_mu} displays the NTRs for four possible discrete
states of $(\mu_{t,1},\mu_{t,2})$ at the initial time. The previous
examples without stochastic parameters have only one unique NTR at
each trading time, but in this example each discrete state of stochastic
parameters has its own corresponding NTR. These NTRs are close to
being square as the two risky assets are i.i.d.. The top-right square
is the NTR for the state $(\mu_{t,1},\mu_{t,2})=(0.08,0.08)$, the
bottom-left square is the NTR for the state $(\mu_{t,1},\mu_{t,2})=(0.06,0.06)$,
and the top-left and the bottom-right squares are the NTRs for the
states $(\mu_{t,1},\mu_{t,2})=(0.06,0.08)$ and $(\mu_{t,1},\mu_{t,2})=(0.08,0.06)$
respectively. The diamond, the mark, the plus, and the circle inside
the squares are the corresponding Merton points if we assume that
$(\mu_{t,1},\mu_{t,2})$ are fixed at their initial values. We see
that a smaller $\mu_{t,i}$ implies a NTR and a Merton point closer
to the origin of the coordinate system. This is consistent with the
formula of the Merton point, (\ref{eq:Merton_point}). We see that
these Merton points are no longer in the centers of the NTRs, like
what the previous examples without stochastic parameters have shown
(except the short-horizon problems); thus using Merton points as portfolio
solution is no longer good as the difference between the optimal solution
and the Merton points may be more than 0.1, as shown in Figure \ref{fig:StoPara_mu}.
Appendix \ref{sec:Problems-with-Stochastic} shows more examples with
stochastic interest rate or volatility, and their NTRs have similar
patterns.

\subsubsection{Example 4: a Four-Asset Problem with Consumption}

\begin{figure}
\begin{centering}
\includegraphics[width=0.5\textwidth]{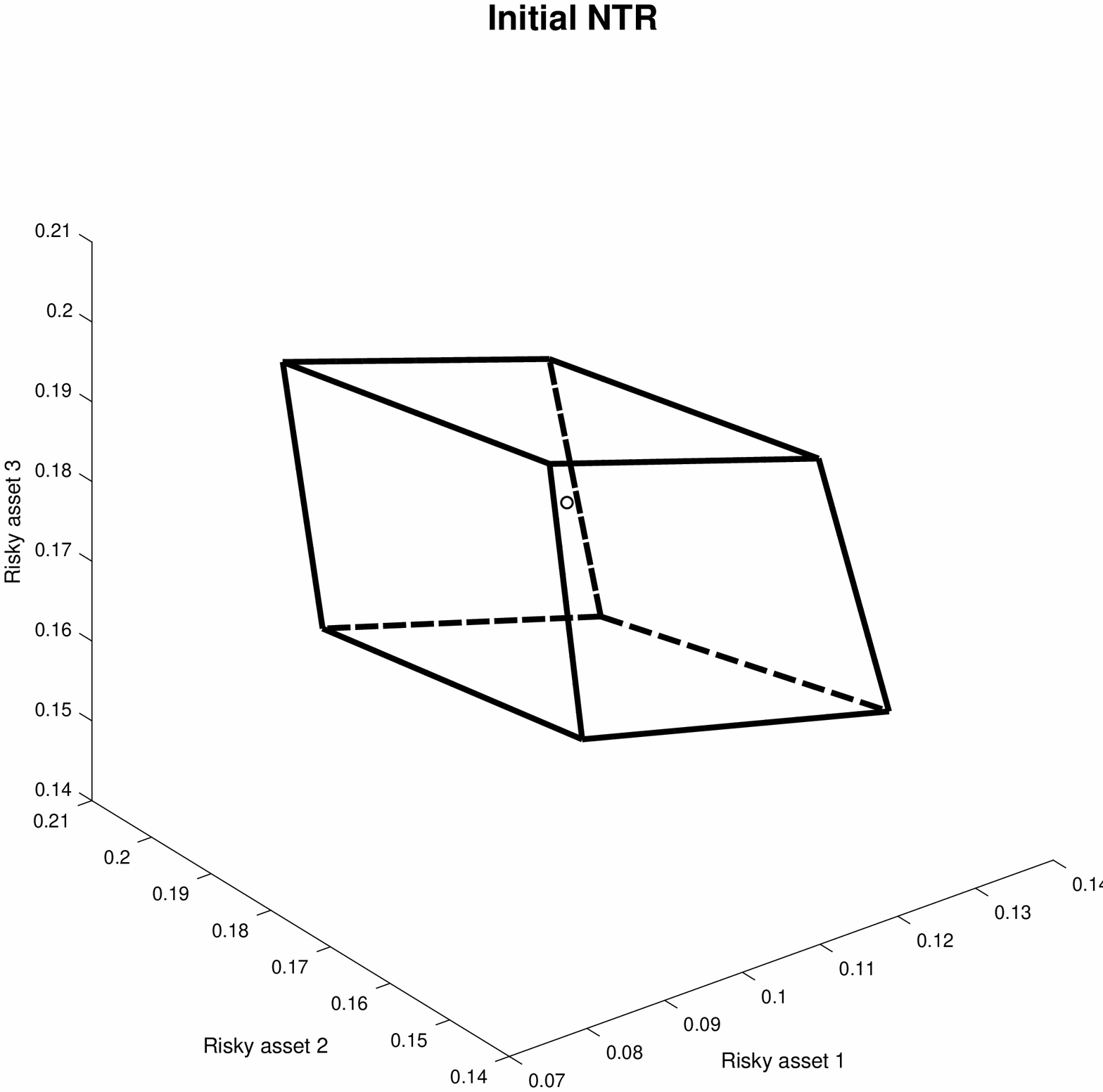}
\par\end{centering}
\caption{\label{fig:Cons3S_0.4}Initial NTR for 3 correlated risky assets and
1 risk-free asset with consumption.}
\end{figure}

It is often challenging for a PDE solution method to solve problems
with three or more continuous state dimensions. In this example, we
employ the numerical DP method to solve the with-consumption model
(\ref{Eq:Port_Power_Cons_Model}) with three continuous state dimensions,
in which we have one risk-free asset and three correlation risky assets
available for monthly trading (i.e., $\Delta t=1/12$ years) in $T=3$
years (so there are $N=36$ periods). The parameter values are listed
in Table \ref{table:parameters_cons}. In the numerical DP method
for this example, we choose degree-$60$ complete Chebyshev polynomials
to approximate value functions, use $61^{3}$ tensor Chebyshev nodes
as the approximation nodes to compute Chebyshev coefficients, and
implement the multi-dimensional product Gauss-Hermite quadrature rule
with $3^{3}$ tensor quadrature nodes. We run the parallel DP method
on the Blue Waters supercomputer, and it takes nearly 2.6 wall clock
hours using 640 cores. A higher degree (e.g., degree-80) polynomial
approximation (with a higher number of Chebyshev nodes) or more quadrature
nodes (e.g., 5 nodes in each dimension) have little change to the
solution: e.g., the $\mathcal{L}^{1}$ difference between the solution
from the degree-60 polynomial approximation and the solution from
the degree-80 polynomial approximation is $4.7\times10^{-4}$, and
the $\mathcal{L}^{\infty}$ difference is 0.0017. Figure \ref{fig:Cons3S_0.4}
displays the NTR at the initial time. The circle inside the NTR is
the Merton point: $(0.1071,0.1786,0.1786)$. For convenience, we plot
the faces of the NTR as flat, but in fact there is small curvature
on the faces, and the exact NTR might have curvy faces too as Figure
\ref{fig:Kumar_NTR} has shown. The NTR is tilted as the risky assets
are correlated.

\subsubsection{Error Analysis for Example 4}

Examples 2 and 4 have shown that computational time expands exponentially
with the number of risky assets as we use tensor grids for approximation
nodes and quadrature nodes. In particular, due to the existence of
kinks of the value functions, we have to use high-degree polynomial
approximations. In Example 4, we have used the degree-80 solution
as the ``true'' solution to compare with lower-degree solutions.
If the difference is small, we consider the solutions from the lower-degree
approximations to be accurate.

However, for problems with more risky assets, we cannot use a solution
from a very high degree polynomial approximation to estimate our solution's
accuracy, due to computational resource limitations. Instead, we roughly
estimate our solution's accuracy from policy approximation errors.
That is, we use the solutions of optimal decisions over all approximation
nodes in the state space to construct a complete Chebyshev polynomial
to approximate a policy function for each decision variable, and then
we compute the distances between optimal decisions over 1,000 random
nodes in the state space and the values of the approximated policy
functions over the same set of random nodes. These distances are called
policy approximation errors. This method has been implemented in Cai
et al. (2017) and Cai and Lontzek (2019).

Table \ref{tab:Error-Analysis-and} shows the policy approximation
errors and the difference from the degree-80 solution in their $\mathcal{L}^{1}$
and $\mathcal{L}^{\infty}$ norms. We see that both the policy approximation
errors and the difference from the degree-80 solution are declining
over the degrees. A lower-degree approximation is computationally
faster but its error is also larger. For example, using degree-8 approximation
takes only 18 seconds using 6 cores, but its policy approximation
error in $\mathcal{L}^{\infty}$ is 0.02, and its solution's difference
from the degree-80 solution is 0.045 in the $\mathcal{L}^{\infty}$
measurement. If we want to control the $\mathcal{L}^{\infty}$ error
at around 0.005, then the degree-30 solution can suffice.

\begin{table}

\begin{centering}
\begin{tabular}{c|c|c|c|c|c|c}
\hline 
Deg. & \multicolumn{2}{c|}{Policy Approx. Errors} & \multicolumn{2}{c|}{Diff. from deg-80 Sol.} & Number & Time\tabularnewline
\cline{2-5} \cline{3-5} \cline{4-5} \cline{5-5} 
 & $\mathcal{L}^{1}$ Error & $\mathcal{L}^{\infty}$ Error & $\mathcal{L}^{1}$ meas. & $\mathcal{L}^{\infty}$ meas. & of Cores & \tabularnewline
\hline 
8 & 0.00190 & 0.0203 & 0.0195 & 0.0452 & 6 & 18 seconds\tabularnewline
20 & 0.00040 & 0.0089 & 0.0059 & 0.0152 & 6 & 12 minutes\tabularnewline
30 & 0.00016 & 0.0062 & 0.0027 & 0.0050 & 256 & 7 minutes\tabularnewline
40 & 0.00013 & 0.0042 & 0.0027 & 0.0046 & 288 & 37 minutes\tabularnewline
60 & 0.00009 & 0.0027 & 0.0005 & 0.0017 & 640 & 2.6 hours\tabularnewline
\hline 
\end{tabular}
\par\end{centering}
\caption{Error Analysis and Running Times.\label{tab:Error-Analysis-and}}

\end{table}

\subsubsection{Example 5: a Five-Asset Problem with Consumption}

Our last example without options solves the with-consumption model
(\ref{Eq:Port_Power_Cons_Model}), where the assets available for
trading include one risk-free asset and four risky assets with independent
log-normal annual returns. We assume $T=3$ years and $\Delta t=1/12$
years. The parameter values are listed in Table \ref{table:parameters_cons}.

\begin{figure}
\begin{centering}
\includegraphics[width=0.5\textwidth]{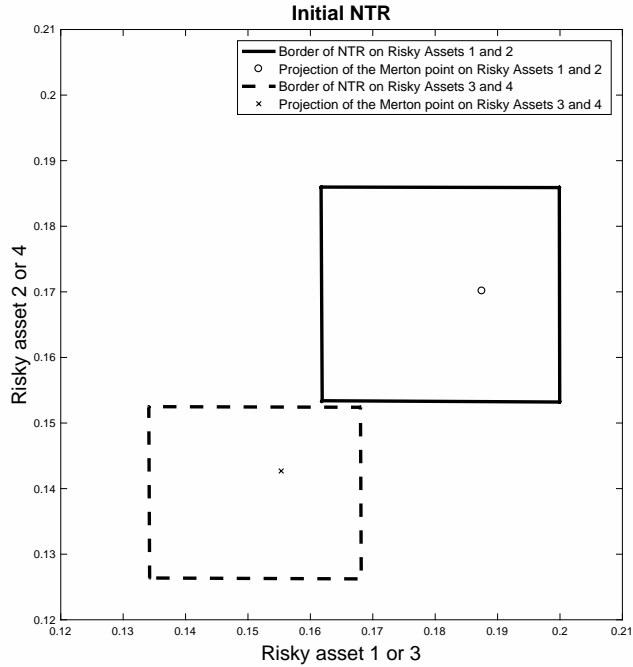}
\par\end{centering}
\centering{}\caption{\label{fig:4S} Initial NTR for 4 independent risky assets and 1 risk-free
asset.}
\end{figure}

Figure \ref{fig:4S} displays the NTR at the initial time, and we
see that the NTR is close to being a hyper-rectangle as the four risky
assets are uncorrelated. The top-right rectangle (with solid lines)
and the bottom-left rectangle (with dashed lines) are the cross-sections
of the NTR for risky assets 1 and 2, and risky assets 3 and 4, respectively.
In the figure, the circle and the mark are, respectively, the projections
of the Merton point, $(0.1875,0.1701,0.155,0.1427)$, on risky assets
1 and 2, and risky assets 3 and 4.

We choose degree-30 complete Chebyshev polynomials to approximate
value functions, use $31^{4}$ tensor Chebyshev nodes to compute Chebyshev
coefficients, and implement the multi-dimensional product Gauss-Hermite
quadrature rule with $3^{4}$ tensor quadrature nodes. Thus, for each
period, there are $31^{4}=923,521$ optimization problems in the maximization
step. Moreover, the evaluation of the objective function of each optimization
problem is time-consuming: the product quadrature rule means there
are $3^{4}=81$ evaluations (one evaluation per quadrature node) of
a degree-30 complete Chebyshev polynomial which contains $\left(\begin{array}{c}
30+4\\
4
\end{array}\right)=46,376$ terms of Chebyshev basis polynomials. This is a computationally intensive
problem. It takes about 9.8 hours to get the solution using 3,840
cores of the Blue Waters supercomputer. If we use degree-60 complete
Chebyshev polynomials with $61^{4}$ tensor Chebyshev nodes, then
we can improve our solution's accuracy, but it would require too much
computing resources: around 
\[
\frac{61^{4}\times\left(\begin{array}{c}
60+4\\
4
\end{array}\right)}{31^{4}\times\left(\begin{array}{c}
30+4\\
4
\end{array}\right)}\approx205
\]
times more as a rough estimate, which has exceeded our available resources
in the Blue Waters supercomputer. Thus, we cannot use a solution from
a very high degree polynomial approximation to estimate our solution's
accuracy, so instead we roughly estimate it using policy approximation
errors, which are $2.3\times10^{-4}$ in $\mathcal{L}^{1}$ and $0.0056$
in $\mathcal{L}^{\infty}$. These errors are close to the policy approximation
errors in Example 4 with the degree-30 approximation. Thus, from Table
\ref{tab:Error-Analysis-and}, we estimate that the error of our solution
in Figure \ref{fig:4S} in comparison with the true solution would
be around 0.003 in $\mathcal{L}^{1}$ and 0.005 in $\mathcal{L}^{\infty}$.

We also run an example with one risk-free asset and six uncorrelated
risky assets using degree-8 complete Chebyshev polynomials to approximate
value functions on the six-dimensional hypercube, and get the NTR
as a six-dimensional hyper-rectangle. Its policy approximation errors
are also close to those of Example 4 with the degree-8 approximation,
so its error would be several percent of wealth, such that it might
not be better than a solution using its Merton point for a long-horizon
problem. However, for such a high-dimensional problem, our numerical
DP method can provide a NTR which might still be better than a Merton
point, when the problem has a short horizon or stochastic parameters,
in which the distance between the Merton point and the true NTR could
be relatively large, as shown in the right panel of Figure \ref{fig:Kumar_NTR}
or Figure \ref{fig:StoPara_mu}.

\section{Pricing Formula for Options\label{sec:Pricing-Formulas-for}}

We first give the pricing formula for a put option. Let the expiration
time of the put option be $T$, and let its strike price be $K$.
Then the payoff of the put option at the expiration time $T$ is $\max(K-S_{T},0)$,
where $S_{T}$ is the price of the underlying risky asset at time
$T$. We assume that the price $S_{t}$ of the underlying asset follows
a geometric Brownian motion process with constant drift $\mu$ and
volatility $\sigma$, and the risk-free interest rate is $r$. Black
and Scholes (1973) give an explicit risk-neutral pricing formula for
the European-type put option at time $t<T$. However, some other kinds
of options cannot have an explicit formula. For example, an American-type
put option has to use some numerical methods for its pricing. One
general numerical method for pricing is the binomial lattice method
(see e.g., Luenberger 1997), which can be used for pricing both American-type
and European-type options.

In the binomial lattice model, we set $h$ as a sub-period length
and $n$ as the number of sub-periods such that $nh=\Delta t$, where
$\Delta t$ is the length of one period in the multi-period portfolio
optimization model (i.e., investors can rebalance their portfolio
every time $\Delta t$). If the price is known as $S$ at the beginning
of a sub-period, the price at the beginning of the next sub-period
is $Su$ or $Sd$, with $u>1$ and $0<d<1$. The probability of the
upward movement is $p$, and the probability of the downward movement
is $1-p$. To match the log-normal assumption of the risky asset return,
we follow the binomial lattice method to assume that 
\begin{equation}
\left\{ \begin{array}{l}
p=\frac{1}{2}+\frac{\mu-\sigma^{2}/2}{2\sigma}\sqrt{h},\\
u=\exp(\sigma\sqrt{h}),\\
d=\exp(-\sigma\sqrt{h}),
\end{array}\right.\label{eq:binomial_lattice}
\end{equation}
such that the expected growth rate of $\log(S)$ in the binomial model
converges to $(\mu-\sigma^{2}/2)$, and the variance of the rate converges
to $\sigma^{2}$, as $h$ goes to zero. The risk-free return of one
sub-period is $\exp(rh)$, and the risk-free return of one period
is $R_{f}=(\exp(rh))^{n}=\exp(r\Delta t)$. Thus, the value of one
sub-period put option over the underlying risky asset governed by
the binomial lattice is given as 
\[
P=\exp(-rh)(qP_{u}+(1-q)P_{d}),
\]
where $P_{u}$ and $P_{d}$ are the values at the upward branch and
the downward branch at the end of this sub-period respectively, and
$q$ is the risk-neutral probability with 
\[
q=(\exp(rh)-d)/(u-d).
\]
Since the payoff of the put option at the expiration time $T$ is
$P_{T}=\max(K-S_{T},0)$, we can compute the price of the put option
at any binomial node by applying the above backward iterative relation.
From the above binomial model, we know that the risky asset return
of one period, $R$, has the following probability distribution: 
\begin{equation}
\Pr(R=u^{j}d^{n-j})=\frac{n!}{j!(n-j)!}p^{j}(1-p)^{n-j}\label{eq:binomial-R}
\end{equation}
for $j=0,1,\ldots,n,$ where $p,u$, and $d$ are defined in (\ref{eq:binomial_lattice}).

We also use the same binomial lattice method to price a call option.
The computational process is the same as for the put option, except
we need to set the payoff of the call option at the expiration time
$T$ as $P_{T}=\max(S_{T}-K,0)$ for a pre-specified strike price
$K$. The binomial lattice method will also be used to price a butterfly
option, a combination of a call option and a put option with the same
strike price. That is, the payoff of the butterfly option at the expiration
time is $P_{T}=\left|S_{T}-K\right|$.

\section{DP Models with Options\label{sec:DP-Models-option}}

We solve dynamic portfolio problems with a portfolio including a risk-less
asset, an option, and its underlying risky asset. We assume that the
option has an expiration time $T$ and a strike price $K$, and it
is available for trading at times $t=t_{0},t_{1},t_{2},\ldots,t_{N}$
with $t_{j}=j\Delta t$ for $j=0,1,...,N$ and $\Delta t=T/N$, with
a price process $P_{t}$, while its underlying risky asset's price
process is $S_{t}$. Since the option price $P_{t}$ is dependent
on $S_{t}$ and $K$, we denote it as $P_{t}(S_{t},K)$ explicitly.
Right before reallocation time $t$, assume that $W_{t}$ is the wealth,
$X_{t}$ is the amount of money invested in the risky asset, and $Y_{t}$
is the amount of money invested in the option. Since the price of
the option is dependent on the risky asset price, we should add $S_{t}$
as a state variable. That is, there are four state variables: $W_{t},X_{t},Y_{t},S_{t}$.
Denote $A_{t}=S_{t}/K$. Since the payoff of a put option is $P_{T}=\max(K-S_{T},0)=K\cdot\max(1-A_{T},0)$,
we have $P_{t}(S_{t},K)=KP_{t}(A_{t},1)$ by replication. For simplicity,
we denote $P_{t}(A_{t})=P_{t}(A_{t},1)$, so 
\[
\frac{P_{t+\Delta t}(S_{t+\Delta t},K)}{P_{t}(S_{t},K)}=\frac{P_{t+\Delta t}(A_{t+\Delta t})}{P_{t}(A_{t})}.
\]
Similarly, for a call option or a butterfly option, the separation
of $K$ and $A_{t}$ still hold. Thus, we can use $A_{t}$ as the
state variable instead of $S_{t}$.

Assume that $\Delta X_{t}$ is the amount of dollars for buying or
selling the risky asset at time $t$, and $\Delta Y_{t}$ is the amount
of dollars for buying or selling the option. $\Delta X_{t}>0$ or
$\Delta Y_{t}>0$ means buying the risky asset or the option, while
$\Delta X_{t}<0$ and $\Delta Y_{t}<0$ means selling the risky asset
and the option. We assume that there are proportional transaction
costs in buying or selling the risky asset and the option, with the
proportions $\tau_{1}$ and $\tau_{2}$ respectively. Thus, the transition
laws of the four state variables in one period with time length $\Delta t$
are

\begin{eqnarray}
A_{t+\Delta t} & = & A_{t}R,\label{eq:next_S}\\
Y_{t+\Delta t} & = & (Y_{t}+\Delta Y_{t})\frac{P_{t+\Delta t}(A_{t+\Delta t})}{P_{t}(A_{t})},\label{eq:next-Y}\\
X_{t+\Delta t} & = & R(X_{t}+\Delta X_{t}),\label{eq:next-X}\\
W_{t+\Delta t} & = & R_{f}(W_{t}-X_{t}-Y_{t}-Z_{t})+X_{t+\Delta t}+Y_{t+\Delta t},\label{eq:next-W}
\end{eqnarray}
where $R$ is the random return of the risky asset in one period with
length $\Delta t$, $R_{f}=\exp(r\Delta t)$ is the one-period risk-free
return with $r$ the interest rate, and 
\begin{equation}
Z_{t}=\Delta X_{t}+\Delta Y_{t}+\tau_{1}\left|\Delta X_{t}\right|+\tau_{2}\left|\Delta Y_{t}\right|.\label{eq:M-def}
\end{equation}

We also assume that there are constraints in shorting risky assets/options
or borrowing cash. For simplicity, we assume that neither shorting
nor borrowing is allowed. That is, we have the following constraints:
\begin{eqnarray}
X_{t}+\Delta X_{t} & \geq & 0,\label{eq:no-short-stock}\\
Y_{t}+\Delta Y_{t} & \geq & 0,\label{eq:no-short-option}\\
X_{t}+Y_{t}+Z_{t} & \leq & W_{t}.\label{eq:no-borrow}
\end{eqnarray}
Thus, the DP model is 
\begin{equation}
V_{t}(W_{t},X_{t},Y_{t},A_{t})=\max_{\Delta X_{t},\Delta Y_{t}}\text{ }\mathbb{E}_{t}\left\{ V_{t+\Delta t}(W_{t+\Delta t},X_{t+\Delta t},Y_{t+\Delta t},A_{t+\Delta t})\right\} ,\label{eq:DP-model}
\end{equation}
subject to the constraints (\ref{eq:next_S})-(\ref{eq:no-borrow}),
for $t=t_{0},t_{1},t_{2},\ldots,t_{N-1}$. Here, $\mathbb{E}_{t}$
is the expectation operator conditional on the information at time
$t$, and the terminal value function is $V_{T}(W,X,Y,S)=U(W)$ for
some given utility function $U$.

\subsection{Problems without Consumption}

In order to use a smooth optimization solver for solving the DP model,
we want to cancel out the absolute operator in the constraint (\ref{eq:M-def}).
This can be done by letting $\Delta X_{t}=W_{t}(\delta_{tx}^{+}-\delta_{tx}^{-})$
and $\Delta Y_{t}=W_{t}(\delta_{ty}^{+}-\delta_{ty}^{-})$, with $\delta_{tx}^{+},\delta_{ty}^{-}\geq0$
such that $\left|\Delta X_{t}\right|$ and $\left|\Delta Y_{t}\right|$
can be substituted by $W_{t}(\delta_{tx}^{+}+\delta_{tx}^{-})$ and
$W_{t}(\delta_{ty}^{+}+\delta_{ty}^{-})$ respectively in the optimization
problems.

If the utility function is CRRA with relative risk aversion coefficient
$\gamma>0$, then we let $x_{t}=X_{t}/W_{t}$ and $y_{t}=Y_{t}/W_{t}$.
By using $W_{t},x_{t},y_{t}$, and $A_{t}$ as state variables, we
can separate $W_{t}$ and $(x_{t},y_{t},A_{t})$ in the value function
$V_{t}(W_{t},x_{t},y_{t},A_{t})$. When $U(W)=W^{1-\gamma}/(1-\gamma)$
with $\gamma>0$ and $\gamma\neq1$, we have 
\begin{equation}
V_{t}(W_{t},x_{t},y_{t},A_{t})=W_{t}^{1-\gamma}\cdot g_{t}(x_{t},y_{t},A_{t}),\label{eq:CRRA-Value-Separate}
\end{equation}
where 
\begin{equation}
g_{t}(x_{t},y_{t},A_{t})=\max_{\delta_{tx}^{+},\delta_{tx}^{-},\delta_{ty}^{+},\delta_{ty}^{-}\geq0}\text{ }\mathbb{E}_{t}\left\{ \Pi_{t+\Delta t}^{1-\gamma}\cdot g_{t+\Delta t}(x_{t+\Delta t},y_{t+\Delta t},A_{t+\Delta t})\right\} ,\label{eq:CRRA-vfi}
\end{equation}
subject to 
\begin{eqnarray}
A_{t+\Delta t} & = & A_{t}R,\label{eq:c1}\\
\xi_{t+\Delta t} & = & R(x_{t}+\delta_{tx}^{+}-\delta_{tx}^{-}),\label{eq:c2}\\
\eta_{t+\Delta t} & = & (y_{t}+\delta_{ty}^{+}-\delta_{ty}^{-})\frac{P_{t+\Delta t}(A_{t+\Delta t})}{P_{t}(A_{t})},\label{eq:c3}\\
z_{t} & = & \delta_{tx}^{+}-\delta_{tx}^{-}+\delta_{ty}^{+}-\delta_{ty}^{-}+\tau_{1}(\delta_{tx}^{+}+\delta_{tx}^{-})+\tau_{2}(\delta_{ty}^{+}+\delta_{ty}^{-}),\label{eq:c4}\\
\Pi_{t+\Delta t} & = & R_{f}(1-x_{t}-y_{t}-z_{t})+\xi_{t+\Delta t}+\eta_{t+\Delta t},\label{eq:c5}\\
x_{t+\Delta t} & = & \xi_{t+\Delta t}/\Pi_{t+\Delta t},\label{eq:c6}\\
y_{t+\Delta t} & = & \eta_{t+\Delta t}/\Pi_{t+\Delta t},\label{eq:c7}
\end{eqnarray}
and the following no-shorting and no-borrowing constraints 
\begin{eqnarray}
x_{t}+\delta_{tx}^{+}-\delta_{tx}^{-} & \geq & 0,\label{eq:no-short-stock-1}\\
y_{t}+\delta_{ty}^{+}-\delta_{ty}^{-} & \geq & 0,\label{eq:no-short-option-1}\\
x_{t}+y_{t}+z_{t} & \leq & 1.\label{eq:no-borrow-1}
\end{eqnarray}
The terminal function is $g_{T}(x,y,A)=1/(1-\gamma)$. Similarly,
when $U(W)=\log(W)$, we can also separate $W_{t}$ and $(x_{t},y_{t},A_{t})$
in the value function $V_{t}(W_{t},x_{t},y_{t},A_{t})$. The approximation
domain of $(x_{t},y_{t})$ can be set as $[0,1]^{2}$ as we do not
allow shorting or borrowing.

Similarly to what we computed in the portfolio problems without options,
we will also compute a no-trade region, $\Omega_{t}$, for the problems
with options, in which $\Omega_{t}$ is defined as 
\[
\Omega_{t}=\{(x_{t},y_{t}):\ (\delta_{tx}^{+})^{\ast}=(\delta_{tx}^{-})^{\ast}=(\delta_{ty}^{+})^{\ast}=(\delta_{ty}^{-})^{\ast}=0\},
\]
where $(\delta_{tx}^{+})^{\ast},(\delta_{ty}^{+})^{\ast}\geq0$ are
optimal fractions of wealth for buying the risky asset and the option,
and $(\delta_{tx}^{-})^{\ast},(\delta_{ty}^{-})^{\ast}\geq0$ are
optimal fractions of wealth for selling the risky asset and the option,
for a given $(x_{t},y_{t},A_{t})$. From the separability of $W_{t}$
and $(x_{t},y_{t},A_{t})$, we see that the optimal portfolio rules
are independent of wealth $W_{t}$. Thus the no-trade regions $\Omega_{t}$
are also independent of $W_{t}$, for CRRA utility functions.

\subsection{Long Horizon Problems with Consumption and Epstein--Zin Preferences\label{sec:Long-Horizon-Problems}}

In some problems, assets are used to finance consumption during the
period $t$. Like what we discussed in Section \ref{subsec:crra},
if the utility function for consumption is $U(C)=C^{1-\gamma}/(1-\gamma)$
and the terminal value function is $V_{T}(W,x,y,A)=W^{1-\gamma}\cdot g_{T}(x,y,A)$
for some given $g_{T}(x,y,A)$, then the Bellman equation (\ref{eq:CRRA-vfi})
changes to 
\begin{equation}
g_{t}(x_{t},y_{t},A_{t})=\max_{c_{t},\delta_{tx}^{+},\delta_{tx}^{-},\delta_{ty}^{+},\delta_{ty}^{-}\geq0}\text{ }U(c_{t})\Delta t+\beta\mathbb{E}_{t}\left\{ \Pi_{t+\Delta t}^{1-\gamma}\cdot g_{t+\Delta t}(x_{t+\Delta t},y_{t+\Delta t},A_{t+\Delta t})\right\} ,\label{eq:CRRA-vfi-cons}
\end{equation}
subject to the constraints (\ref{eq:c1})-(\ref{eq:c4}), (\ref{eq:c6})-(\ref{eq:no-short-option-1}),
and 
\begin{eqnarray*}
 &  & \Pi_{t+\Delta t}=R_{f}(1-x_{t}-y_{t}-z_{t}-c_{t}\Delta t)+\xi_{t+\Delta t}+\eta_{t+\Delta t},\\
 &  & x_{t}+y_{t}+z_{t}+c_{t}\Delta t\leq1,
\end{eqnarray*}
where $\beta=\exp(-\rho\Delta t)$ is the one-period discount factor.
The terminal value function is 
\begin{equation}
g_{T}(x_{T},y_{T},A_{T})=\frac{U(r(1-\tau_{1}x_{T}))\Delta t}{1-\beta}=\frac{(r(1-\tau_{1}x_{T}))^{1-\gamma}\Delta t}{(1-\beta)(1-\gamma)}\label{eq:term-val-cons}
\end{equation}
by assuming that the agent sells all of the underlying risky asset
(incurring proportional transaction cost $\tau_{1}$) and exercises
all the options (without transaction costs) at the terminal time,
and then consumes only the interest of the terminal wealth forever.

In the model (\ref{eq:CRRA-vfi-cons}), $\gamma$ is the risk aversion
parameter, but it is also the inverse of the intertemporal elasticity
of substitution (IES). Epstein and Zin (1989) introduce a recursive
utility function to separate risk aversion and IES. If we use Epstein--Zin
preferences and $U(c)=c^{1-\gamma}/(1-\gamma)$ with $\gamma$ the
inverse of IES, then the Bellman equation (\ref{eq:CRRA-vfi-cons})
becomes 
\begin{equation}
g_{t}(x_{t},y_{t},A_{t})=\max_{c_{t},\delta_{tx}^{+},\delta_{tx}^{-},\delta_{ty}^{+},\delta_{ty}^{-}\geq0}\text{ }U(c_{t})\Delta t+\beta\left[\mathbb{E}_{t}\left\{ \left(\Pi_{t+\Delta t}^{1-\gamma}\cdot g_{t+\Delta t}(x_{t+\Delta t},y_{t+\Delta t},A_{t+\Delta t})\right)^{\frac{1-\psi}{1-\gamma}}\right\} \right]^{\frac{1-\gamma}{1-\psi}},\label{eq:CRRA-vfi-EZ}
\end{equation}
where $\psi$ is the risk aversion parameter. See Cai et al. (2017)
and Cai and Lontzek (2019) for more discussion on Epstein--Zin preferences
and its use in dynamic stochastic optimization problems (e.g., the
Bellman equation (\ref{eq:CRRA-vfi-EZ}) formula holds only when $0<\gamma<1$;
when $\gamma>1$, it needs some adjustment).

In reality, the time to maturity of options is typically not more
than one year. Thus, if we want to solve a multi-year portfolio optimization
problem with options (where each period is equal to or less than 1
year), we need to liquidate the option at its expiration time and
buy some newly issued options at some intermediate stages. For simplicity,
we assume that all options have the same maturity $T$ while the whole
trading horizon is $mT$ (that is, we have $m$ rounds of option issuing),
and at the new option issuing times $t\in{\cal I}\equiv\{0,T,2T,\ldots,(m-1)T\}$,
only the at-the-money options are available (i.e., the strike price
$K$ for an option issued at time $t\in{\cal I}$ is $S_{t}$, so
the strike prices will be different for these different at-the-money
options issued at different times). Thus, when $t\not\in{\cal I}$
or $t=0$, the DP model is the same as (\ref{eq:DP-model}).

When $t\in{\cal I}$ and $t>0$, we use $t^{+}$ to denote the time
right after liquidating the expired options and before rebalancing
with newly issued options. We assume that there is no transaction
cost for liquidating the expired options. Thus, we have $Y_{t^{+}}=0$
and $W_{t^{+}}=W_{t}$ (while the cash amount increases from $W_{t}-X_{t}-Y_{t}$
to $W_{t}-X_{t}$). Moreover, since we assume that only an at-the-money
option is available for trading no matter what the risky asset price
is at time $t\in{\cal I}$, we have $A_{t^{+}}=1$. That is, we have
\[
V_{t}(W_{t},X_{t},Y_{t},A_{t})=V_{t^{+}}(W_{t^{+}},X_{t^{+}},0,1)=V_{t^{+}}(W_{t},X_{t},0,1)=W_{t}^{1-\gamma}\cdot g_{t}(x_{t},0,1),
\]
for $t\in{\cal I}$ and $t>0$, because we assume that there is no
transaction cost for liquidating the expired options.

\section{Examples with Options \label{sec:Examples-option}}

In all examples in this section, we assume that the available assets
for trading are one risk-free asset, one risky asset, and one option
underlying the risky asset. The risk-free asset has an annual interest
rate, $r=0.01$. The risky asset has a return $R$ drawn from the
binomial distribution (\ref{eq:binomial-R}) with $\mu=0.07$, $\sigma=0.2$,
and a proportional transaction cost ratio $\tau_{1}$. The option
has an expiration time $T$ and a strike price $K$ specified at its
issue time and a proportional transaction cost ratio $\tau_{2}$.
The terminal value function is $U(W)=W^{1-\gamma}/(1-\gamma)$ with
the default $\gamma=3$ for the problems without consumption. The
default length of one period is $\Delta t=1/52$ years, i.e., one
week. We apply the binomial lattice model (\ref{eq:binomial_lattice})
with the default length of one sub-period $h=1/520$ (so the default
number of sub-periods is $n=\Delta t/h=10$).

In the following examples, our approximation method in numerical DP
uses degree-100 complete Chebyshev polynomials with $101^{2}$ Chebyshev
nodes in two continuous states: $(x_{t},y_{t})$, where $x_{t}$ and
$y_{t}$ are the fractions of wealth invested in the risky asset and
the option at the time right before the reallocation time $t$ respectively.
Similarly to our examples without options, we use the square $[0,1]^{2}$
as the approximation domain for $(x_{t},y_{t})$, and we choose a
degree-100 polynomial approximation as a higher degree approximation
(with a higher number of approximation nodes) has little change in
the solution. Here we do not need the Gauss-Hermite quadrature rule
as we have already discretized the price of the risky asset and keep
it as a discrete state variable.

\subsection{Portfolio with a Put Option}

\begin{figure}[htb]
\begin{centering}
\begin{tabular}{cc}
\includegraphics[width=0.5\textwidth]{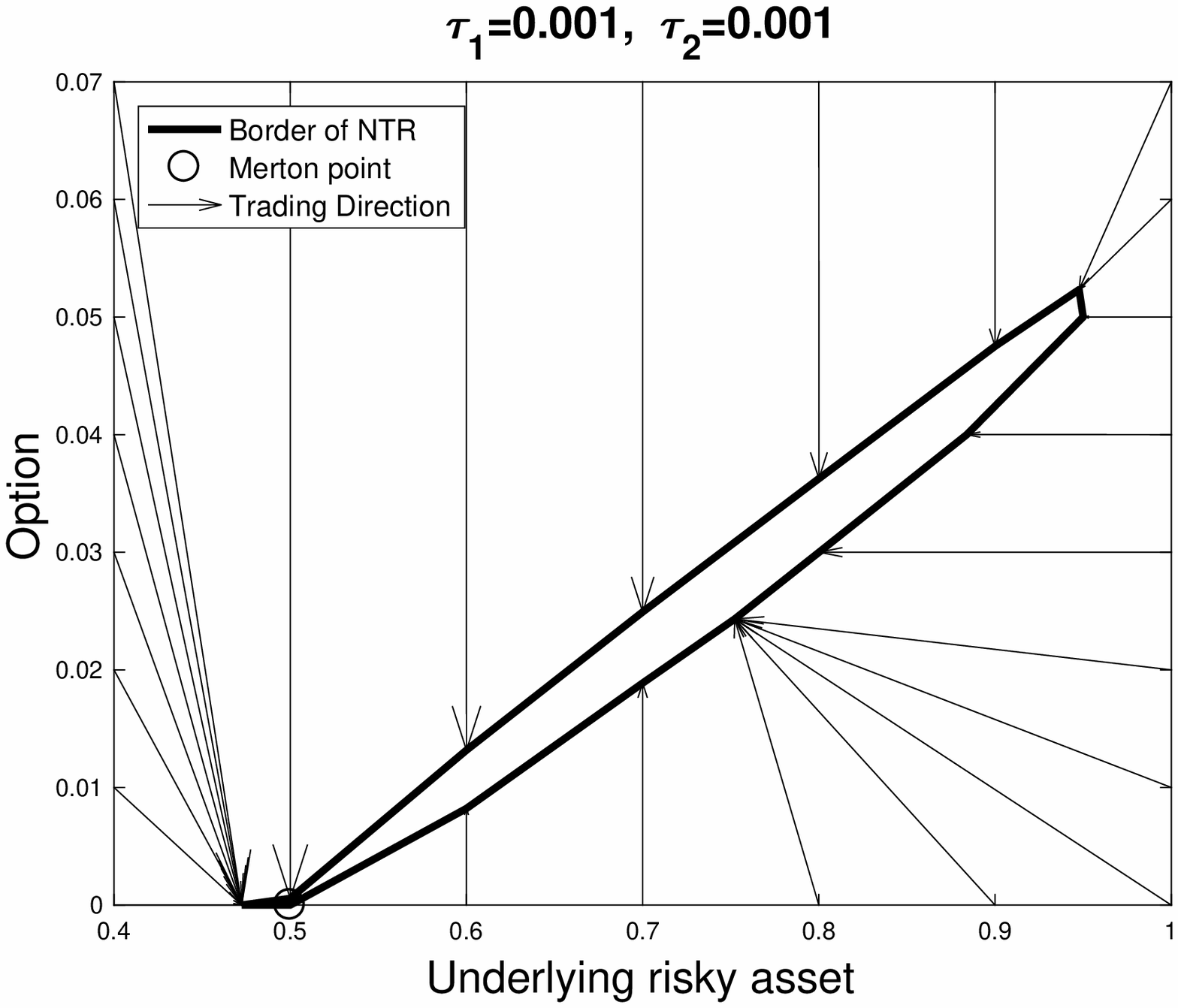} & \includegraphics[width=0.5\textwidth]{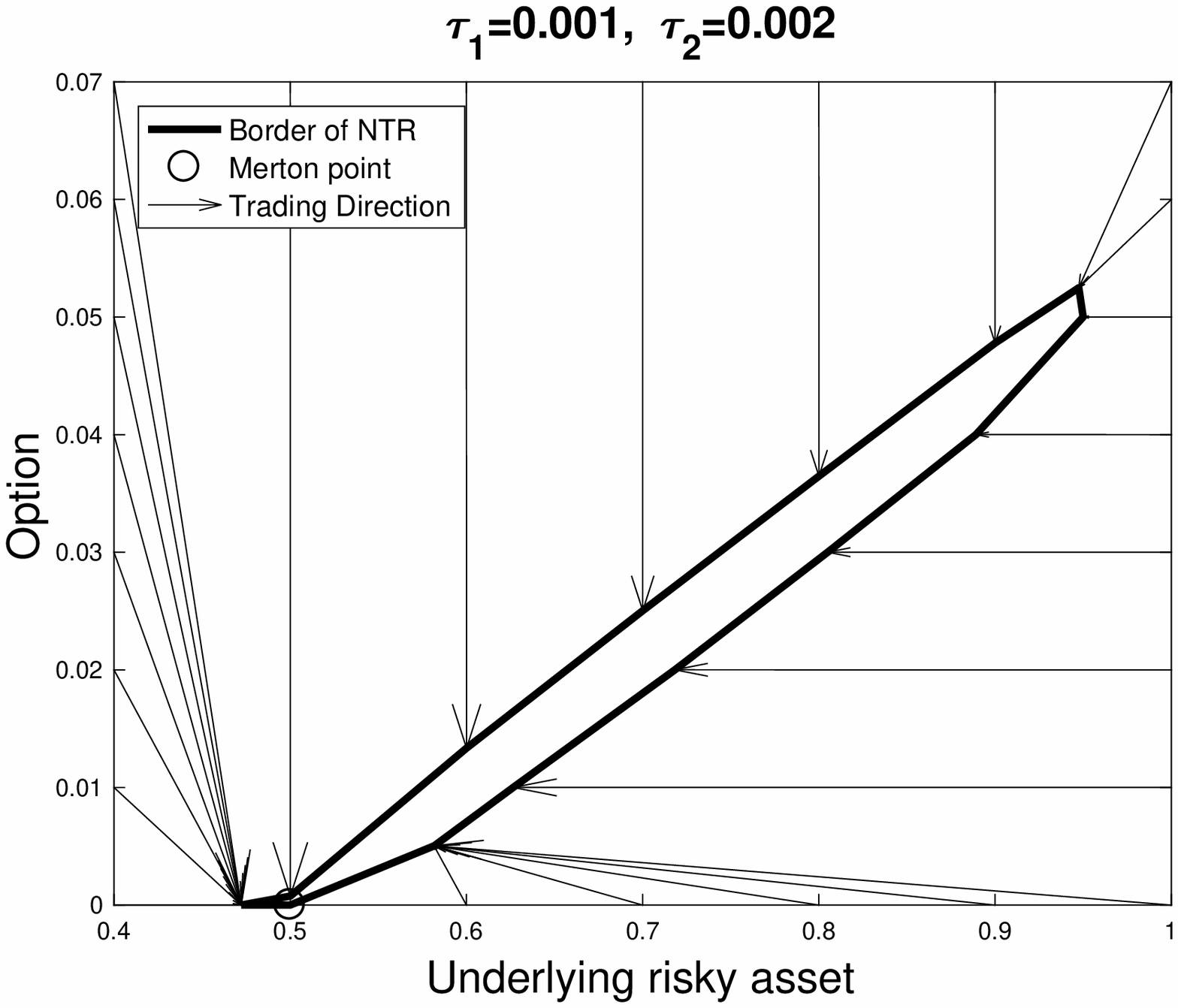}\tabularnewline
\end{tabular}
\par\end{centering}
\centering{}\caption{\label{fig:NTR-put} Initial NTRs for 1 underlying risky asset, 1
put option, and 1 risk-free asset.}
\end{figure}

Our first example with options chooses the at-the-money put option
(i.e., the strike price $K=S_{0}$ or equivalently $A_{0}=1$) with
the expiration time $T=0.5$ years. The trading horizon is also $T$,
so it has 26 re-allocation times with the weekly time steps, and 261
sub-periods over the whole horizon (so the largest number of values
of the discrete state $A_{t}$ is 261, which happens at the terminal
time). Figure \ref{fig:NTR-put} shows the NTRs and the optimal transaction
paths in arrows at the initial time for $\tau_{1}=\tau_{2}=0.001$
(the left panel) and $\tau_{1}=0.001$ and $\tau_{2}=0.002$ (the
right panel), where the horizontal and vertical axes represent the
fraction of wealth invested in the underlying risky asset and the
option respectively (the axes of all other figures in this section
have the same meaning). The circle on the horizontal axis is the Merton
point, $(0.5,0)$, if we assume that the underlying risky asset can
be traded at any time without transaction costs so that the option
can be replicated completely by the risk-free asset and the underlying
risky asset (then there is zero fraction of wealth invested in the
option at the Merton point).

Since the payoff of the put option at the expiration time is $\max(K-S_{T},0)$,
we know that the put option is negatively and highly correlated with
the underlying risky asset. The high correlation of the underlying
risky asset and the option implies that the objective function of
the optimization problem is flat so the NTR should be a strip that
is close to a line, and the strip should be tilted toward one axis.
The negative correlation implies that the strip should be tilted with
a positive slope. All these properties can be observed in Figure \ref{fig:NTR-put},
while the right panel's strip is slightly wider than the left panel's,
due to the larger transaction cost of the option.

The arrows in Figure \ref{fig:NTR-put} show the paths from the initial
allocations on the border of the axis box to the optimal allocations.
In the left panel of Figure \ref{fig:NTR-put} (with $\tau_{1}=\tau_{2}=0.001$),
we see that if the initial allocation of wealth is 47\% or less, then
the optimal decision is to sell all put options and buy some risky
assets until the point (0.47,0). This happens because the put options
are for hedging risks, so if the fraction of risky assets is small,
the whole portfolio's risk is small and it becomes inefficient to
hold the put options. Moreover, if the initial allocation of wealth
is 80\% or more in the underlying risky asset and 0\% in the put option,
then the optimal decision is to sell some of the risky asset until
its fraction becomes 75\% and to buy some put options until its fraction
becomes 2.4\%, in order to hedge risks. Otherwise, holding too many
risky assets may incur too large a loss of wealth if the return of
the risky asset turns out to be small. Furthermore, if the initial
allocation of wealth is between 60\% and 75\% in the underlying risky
asset and 0\% in the put option, then to hedge risks the optimal decision
is to keep the same amount of the risky asset but to buy put options
until the lower envelope of the NTR is reached. However, in the right
panel of Figure \ref{fig:NTR-put} (with $\tau_{1}=0.001$ and $\tau_{2}=0.002$),
if the initial allocation of wealth is 60\% or more in the underlying
risky asset and 0\% in the put option, then the optimal decision is
to sell some of the risky asset until its fraction becomes 58\% and
to buy some put options until its fraction becomes 0.5\%. That is,
the higher transaction cost for options makes an investor less willing
to buy options, and hold less of the risky asset as the need for risk
hedging from put options is also less.

\begin{figure}[htb]
\begin{centering}
\includegraphics[width=0.5\textwidth]{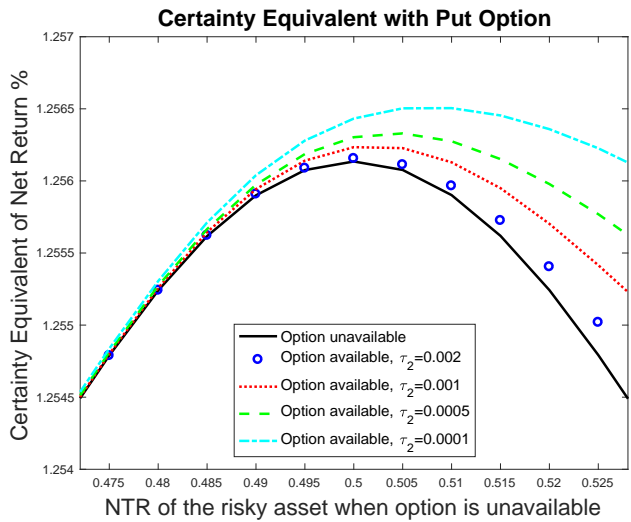}
\par\end{centering}
\centering{}\caption{\label{fig:put-CEQ} Certainty equivalents for put options with various
transaction costs.}
\end{figure}

For each initial risky asset allocation fraction from 0 to 1, while
the initial option allocation fraction is 0, we can compute the optimal
trading strategy for the dynamic portfolio problem, and then we can
compute its certainty equivalent, which is the certain amount giving
the same utility as the expected utility of uncertain future wealth.
Figure \ref{fig:put-CEQ} shows the functions of certainty equivalents
over the initial risky asset allocation fraction for five cases. Its
horizontal axis is the NTR of the underlying risky asset with $\tau_{1}=0.001$
when there is no option available for trading. The solid line represents
the case where there is no option available for trading, and the other
four lines (circled, dotted, dash-dotted, dashed) represent the cases
with different transaction cost ratios, $\tau_{2}$, for the available
option. The figure shows that the put option contributes to the certainty
equivalents. If the allocation fraction in the underlying risky asset
is higher, then the difference is higher, as an investor requires
more put options to hedge risks. Moreover, if $\tau_{2}$ is smaller,
then the corresponding certainty equivalent with the put option is
higher, as the option with smaller transaction costs becomes more
attractive to investors.

From Figure \ref{fig:put-CEQ}, we see that when the initial risky
asset allocation fraction is 52.8\% of the wealth (the upper bound
of the NTR without options), the difference of the certainty equivalent
between the case with the put option with $\tau_{2}=0.001$ and the
case without options is less than 0.001\%. That is, the social value
of one put option is less than 0.00001 dollars per dollar of investment
if initially 52.8\% of total wealth is invested in its underlying
risky asset and the remained amount of the wealth is invested in the
risk-free asset.

\subsection{Portfolio with a Call Option}

\begin{figure}[htb]
\begin{centering}
\begin{tabular}{cc}
\includegraphics[width=0.5\textwidth]{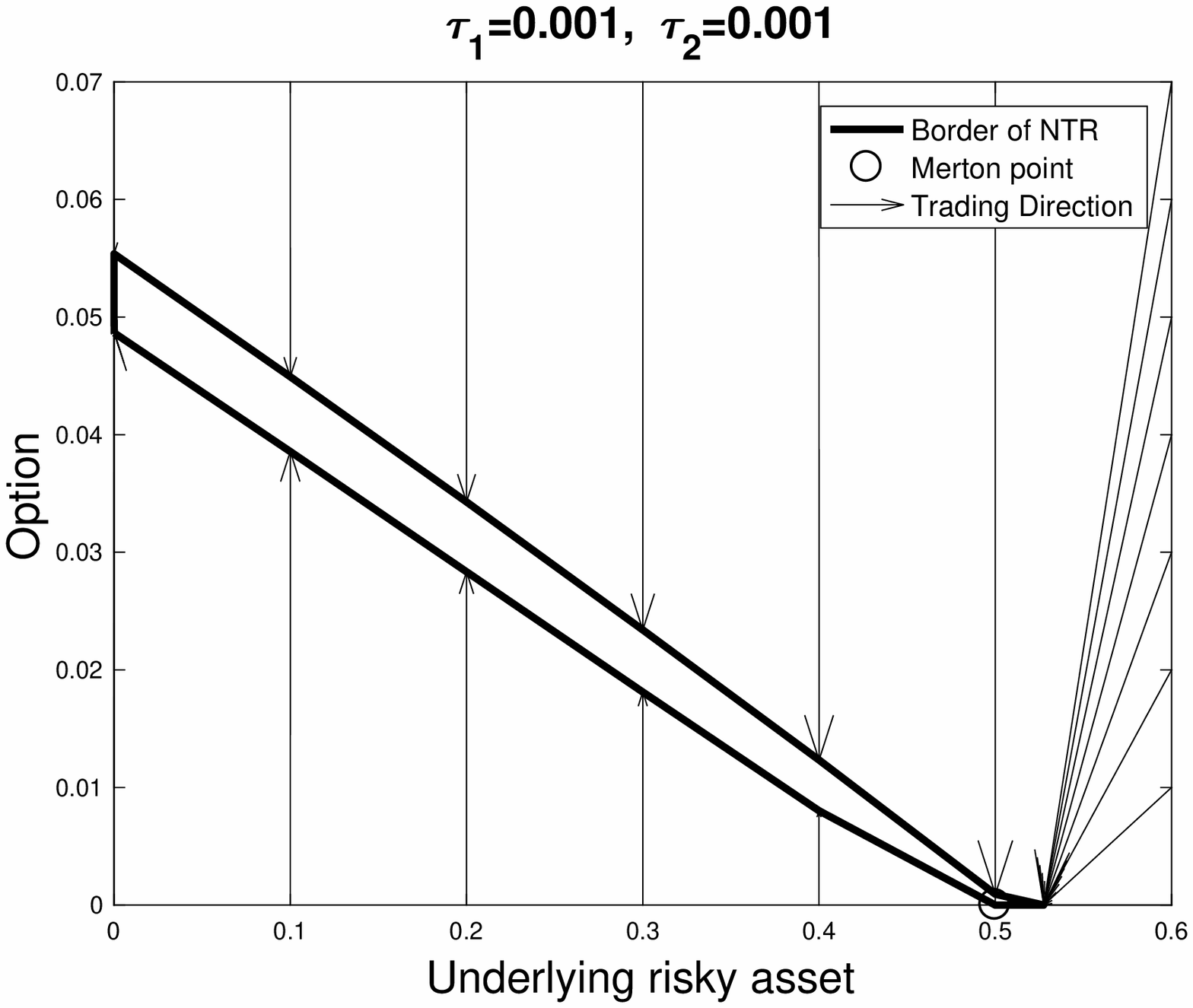}  & \includegraphics[width=0.5\textwidth]{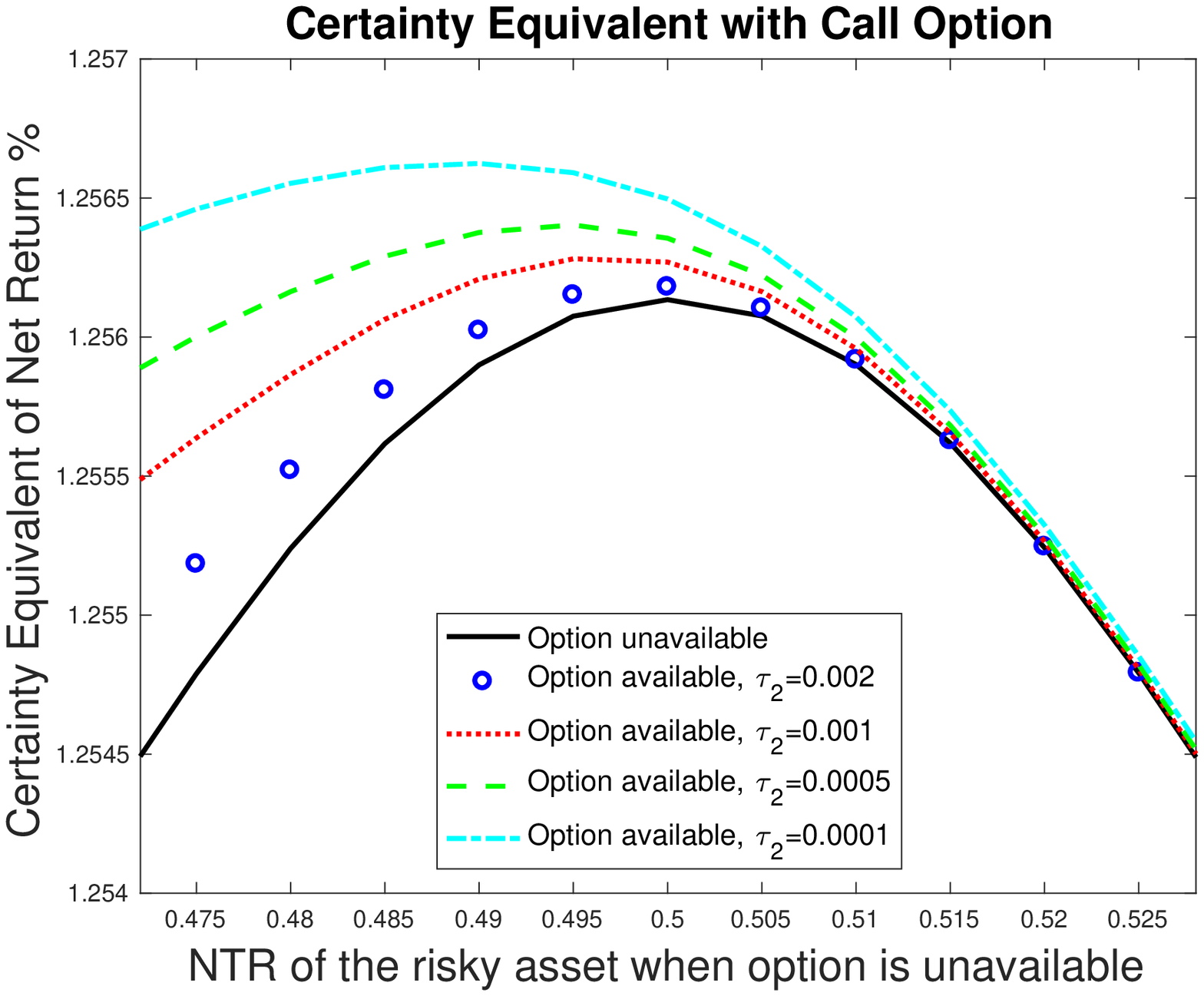}\tabularnewline
\end{tabular}
\par\end{centering}
\centering{}\caption{\label{fig:NTR-call} Initial NTR and certainty equivalents for call
options.}
\end{figure}

Our second example with options uses an at-the-money call option (i.e.,
the strike price $K=S_{0}$ or equivalently $A_{0}=1$) in the portfolio.
Other parameter values are the same as the previous example. The left
panel of Figure \ref{fig:NTR-call} shows the NTR and the optimal
transaction paths in arrows. We see that the NTR is a narrow strip
with a negative slope. This is caused by the positive high correlation
between the call option and its underlying risky asset, as the payoff
of the call at the expiration time is $\max(S_{T}-K,0)$. Moreover,
if the initial allocation fraction in the risky asset is bigger than
0.53, then the optimal decision is to sell all options and to sell
some of the risky asset until the boundary of the NTR (i.e., the point
$(0.53,0)$) is reached. Furthermore, if the initial portfolio holds
the risky asset with 40\% or less of wealth and zero options, then
the optimal decision is to buy the call option until the lower envelope
of the NTR is reached.

The right panel of Figure \ref{fig:NTR-call} shows the functions
of the certainty equivalents of the expected utility of future wealth
over the initial allocation fraction of wealth in the underlying risky
asset for five cases. Similarly to the cases with the put option,
Figure \ref{fig:NTR-call} shows that the call option contributes
to the certainty equivalents, and that the call option is more useful
when the initial allocation in the risky asset is smaller. Moreover,
if the transaction cost ratio for the call option is smaller, then
the corresponding certainty equivalent with the call option is higher,
which is consistent with the intuition that investors like to invest
more in assets with lower transaction costs.

\subsection{Portfolio with a Butterfly Option}

\begin{figure}[H]
\begin{centering}
\begin{tabular}{cc}
\includegraphics[width=0.5\textwidth]{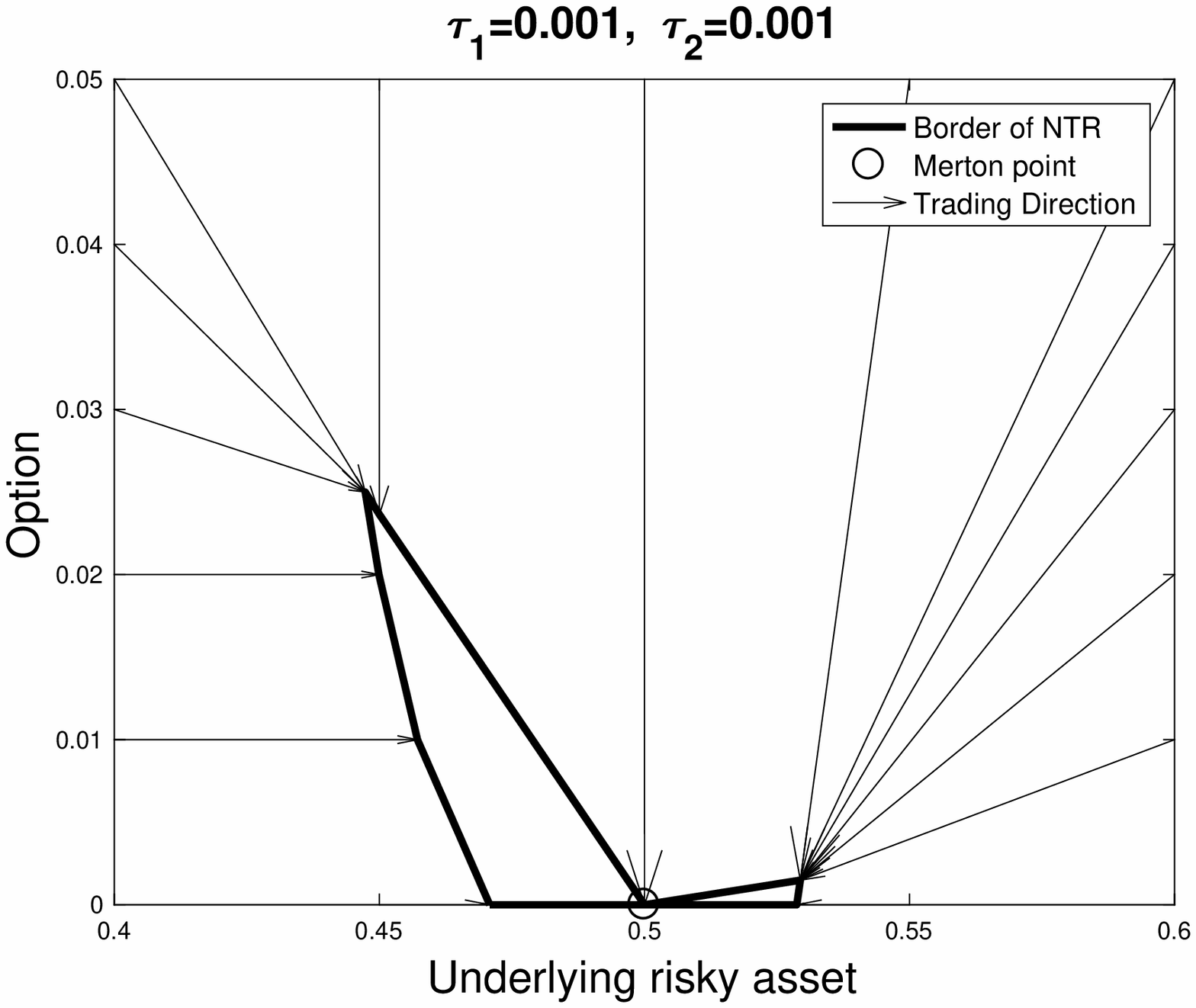}  & \includegraphics[width=0.5\textwidth]{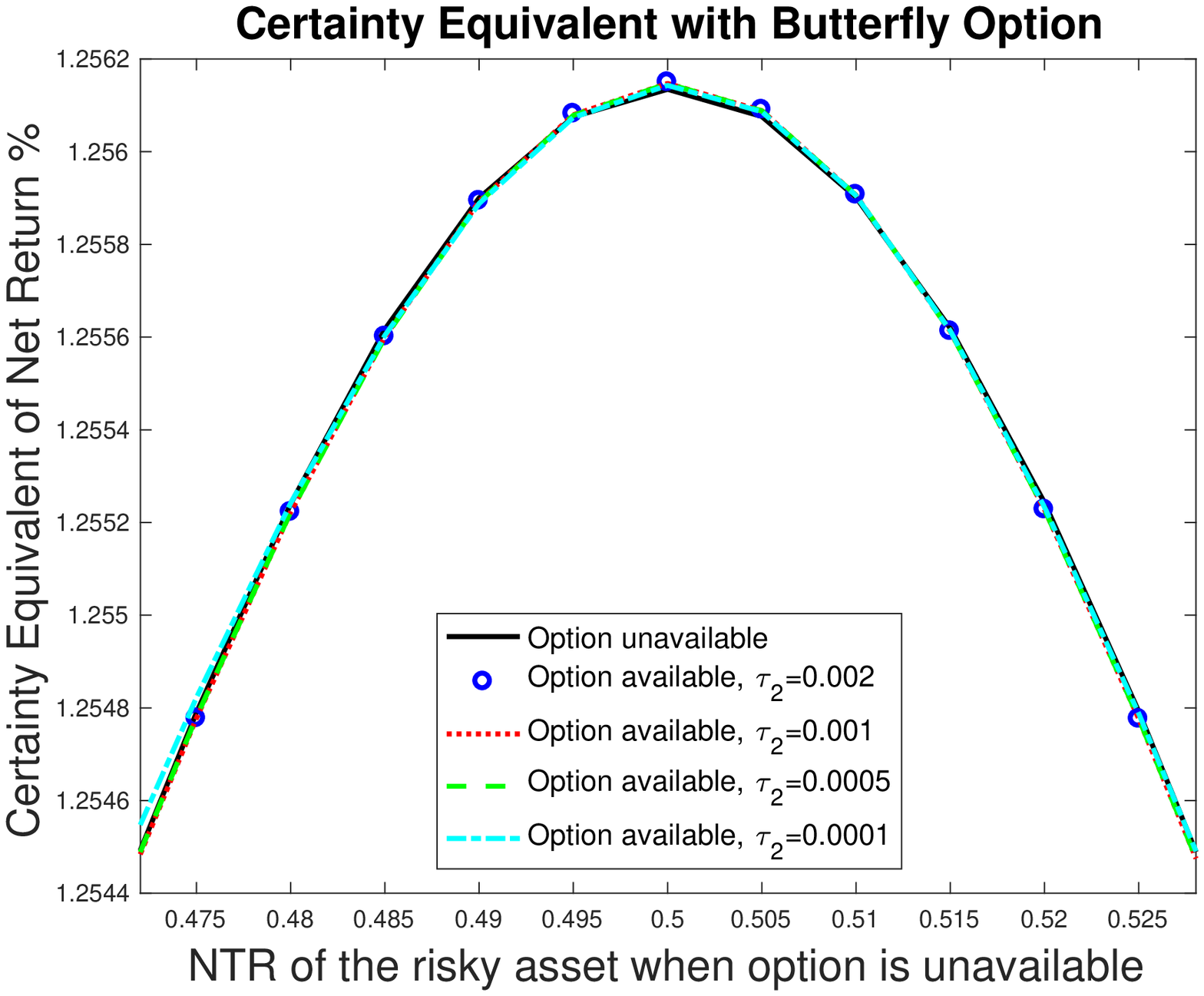}\tabularnewline
\end{tabular}
\par\end{centering}
\centering{}\caption{\label{fig:NTR-butterfly} Initial NTR and certainty equivalents for
butterfly options.}
\end{figure}

Our third example with options uses an at-the-money butterfly option
in the portfolio. Other parameter values are the same as the previous
example. The left panel of Figure \ref{fig:NTR-butterfly} shows the
NTR and the optimal transaction paths in arrows. We see that the NTR
is not a narrow strip like what the previous two examples showed.
Its shape is reasonable because the butterfly option is equivalent
to a combination of an at-the-money put option and an at-the-money
call option, and its payoff at the expiration time is $\left|S_{T}-K\right|$.
Moreover, the optimal decision is to not buy any butterfly option
in the case with $\tau_{1}=\tau_{2}=0.001$. This is also reflected
in the right panel of Figure \ref{fig:NTR-butterfly}, which shows
that the butterfly option has almost no contribution to the certainty
equivalents. This happens because the price of the butterfly is the
sum of the prices of the call and the put, so it becomes less profitable
for investment.

\subsection{Long-Horizon Examples with Consumption and Epstein--Zin Preferences}

We solve the long-horizon model in Section \ref{sec:Long-Horizon-Problems},
which also includes consumption financed by assets, and we use Epstein--Zin
preferences. The trading horizon is three years and $\tau_{1}=\tau_{2}=0.001$.
The expiration time of each newly-issued at-the-money put option is
$T=0.5$ years. The portfolio is rebalanced every week (i.e., $\Delta t=1/52$
years) so it has 156 periods. Every six months there is a newly-issued
at-the-money put option available for trading. That is, at times $t=0.5$,
1, 1.5, 2, 2.5 years, we have a new at-the-money put option available
for buying, while its older put option has to be exercised, so the
largest number of discrete values of $A_{t}$ is 261, which happens
at the maturity times of the option. Following Cai et al. (2017) and
Cai and Lontzek (2019), we choose the utility discount rate to be
$\rho=0.015$ (i.e., the one-period utility discount factor is $\beta=\exp(-\rho\Delta t)$),
$\gamma=2$ or 2/3 (i.e., IES is 0.5 or 1.5), and the degree of risk
aversion (RA) $\psi=2$ or 5. Our numerical DP method employs degree-100
complete Chebyshev polynomials to approximate values functions for
every discrete state. We use 1,600 cores of the Blue Waters supercomputer
and it takes about 1.3 hours for each case of IES and RA.

\begin{figure}[htb]
\begin{centering}
\includegraphics[width=0.5\textwidth]{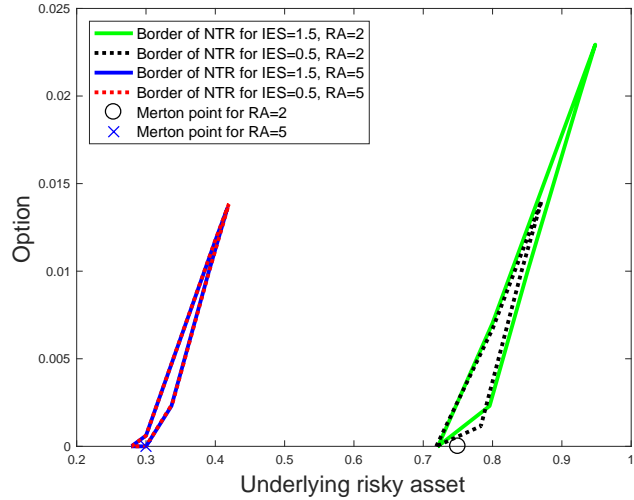}
\par\end{centering}
\centering{}\caption{\label{fig:3y-put-EZ-NTR} NTRs for problems with consumption and
Epstein-Zin preferences.}
\end{figure}

Figure \ref{fig:3y-put-EZ-NTR} shows the NTRs for the four cases
of IES and RA. We see that a more risk averse investor chooses to
hold smaller amounts of risky assets and options, just like what their
Merton points show, so RA does have significant impact on the NTR.
Meanwhile, IES has little impact on the NTR: when RA=5, the difference
between the NTRs when IES=0.5 and IES=1.5 is almost invisible in Figure
\ref{fig:3y-put-EZ-NTR}.

However, IES impacts the optimal consumption rate: when RA=5, the
optimal consumption rate is around 0.012 when IES=0.5, increasing
to around 0.018 when IES=1.5, while the impact of RA on the optimal
consumption is small. That is, a larger IES implies a larger consumption
at the initial time, when the interest rate $r$ is lower than the
utility discount rate $\rho$. This can be explained intuitively by
a simple dynamic programming problem: an investor maximizes $\sum_{j=0}^{\infty}\beta^{j}U(c_{t_{j}})\Delta t$
subject to $W_{t+\Delta t}=R_{f}W_{t}-c_{t}\Delta t$ for any $t=t_{0},t_{1},t_{2},...$,
that is, the only available asset for investment is the risk-free
asset. It has the Euler equation: $U'(c_{t})=\beta R_{f}U'(c_{t+\Delta t})$.
If $U(c)=c^{1-\gamma}/(1-\gamma)$, so its annual consumption growth
rate at the period $(t,t+\Delta t)$ is $g_{t}\equiv\ln(c_{t+\Delta t}/c_{t})/\Delta t=(r-\rho)/\gamma$
, consistent with the famous Ramsey growth theory. That is, if $r<\rho$,
then consumption growth is negative, and a smaller $\gamma$ (i.e.,
a larger IES) implies a larger magnitude of negative growth of consumption,
so then initial consumption is larger. With Epstein-Zin preferences
and risky assets including options, if RA is very large, then almost
all money would be invested in the risk-free asset; this implies that
a larger IES leads to a larger consumption at the initial time. We
also verify this intuition numerically by solving the same long-horizon
examples except with $r$ increased to $0.02$. In the cases with
$r=0.02>\rho=0.015$, we find that a larger IES leads to smaller consumption
at the initial time numerically, as a higher interest rate implies
that future wealth could be much larger if initial consumption is
smaller. This finding is also similar to what Cai et al. (2017) and
Cai and Lontzek (2019) find for the impact of IES and RA on social
cost of carbon.

We also run examples with ten years as the trading horizon while other
parameters (including $r=0.01$) are unchanged. We find that the NTRs
of the ten-year-horizon problems are almost identical to the NTRs
of the three-year-horizon problems at the initial time. 

\section{Conclusion\label{sec:Conclusion}}

This paper has shown that numerical DP can solve multi-stage portfolio
optimization problems with multiple assets and transaction costs in
an efficient and accurate manner. We illustrate trading strategies
by describing the no-trade regions for various choices of asset returns
and transaction costs. The numerical DP algorithms may be computationally
intensive for large portfolio optimization problems, but modern hardware
and parallel DP algorithms make it possible to solve such big problems.
This paper has also shown that when there are transaction costs and
constraints in shorting or borrowing, there exists a thin no-trade
region for the portfolio of an option and its underlying risky asset.
Future research could incorporate sparse grid methods to solve even
larger dynamic portfolio problems.

\newpage{}

\part*{Online Appendices}

\begin{doublespace}
\global\long\def\thefigure{A.\arabic{figure}}%
 \setcounter{figure}{0} 
\global\long\def\thesection{A.\arabic{section}}%
 \setcounter{section}{0} 
\global\long\def\thetable{A.\arabic{table}}%
 \setcounter{table}{0} 
\global\long\def\theequation{A.\arabic{equation}}%
 \setcounter{equation}{0}
\global\long\def\thepage{A.\arabic{page}}%
 \setcounter{page}{1} 
\end{doublespace}

\section{Numerical DP Algorithms\label{sec:Numerical-DP-Algorithms}}

If the state and control variables in a DP problem are continuous,
then the value function must be approximated in some computationally
tractable manner. It is common to approximate value functions with
a finitely parameterized collection of functions; that is, we use
some functional form $\hat{V}(\mathbf{x};{\bf b})$, where ${\bf b}$
is a vector of parameters, and approximate a value function, $V(\mathbf{x})$,
with $\hat{V}(\mathbf{x};{\bf b})$ for some parameter value ${\bf b}$.
For example, $\hat{V}$ could be a linear combination of polynomials
where ${\bf b}$ would be the weights on polynomials. After the functional
form is fixed, we focus on finding the vector of parameters, ${\bf b}$,
such that $\hat{V}(\mathbf{x};{\bf b})$ approximately satisfies the
Bellman equation.

Numerical solutions to a DP problem are based on the Bellman equation:
\begin{eqnarray}
 &  & V_{t}(\mathbf{x},\mathbf{\mathbf{\theta}})=\max_{\mathbf{a}\in{\cal D}(\mathbf{x},\mathbf{\mathbf{\theta}},t)}\text{ \ }U_{t}(\mathbf{x},\mathbf{a})+\beta\mathbb{E}_{t}\left\{ V_{t+\Delta t}(\mathbf{x}^{+},\mathbf{\mathbf{\theta}}^{+})\right\} ,\label{eq:StoDP_Model}\\
 &  & \text{{\rm \ \ \ \ \ \ \ \ \ \ \ \ \ \ \ \ s.t. \ \ }}\mathbf{x}^{+}=\mathbf{\mathcal{G}}_{t}(\mathbf{x},\mathbf{\mathbf{\theta}},\mathbf{a},\mathbf{\omega}),\nonumber \\
 &  & \text{{\rm \ \ \ \ \ \ \ \ \ \ \ \ \ \ \ \ \ \ \ \ \ \ \ }}\mathbf{\mathbf{\theta}}^{+}=\mathbf{\mathcal{H}}_{t}(\mathbf{\mathbf{\theta}},\mathbf{\epsilon}),\nonumber 
\end{eqnarray}
where $\mathbf{x}$ is the vector of continuous states in $\mathbb{R}^{k}$
(in our portfolio examples, $k$ is the number of risky assets including
options), $\mathbf{\mathbf{\theta}}$ is the vector of discrete states
in a finite set $\Theta$, $V_{t}(\mathbf{x},\mathbf{\mathbf{\theta}})$
is called the value function at time $t\leq T$ (the terminal value
function $V_{T}(\mathbf{x},\mathbf{\mathbf{\theta}})$ is given),
$\mathbf{a}$ is the action variable vector in its feasible set ${\cal D}(\mathbf{x},\mathbf{\mathbf{\theta}},t)$,
$\mathbf{x}^{+}$ is the next-stage continuous state vector with its
transition function $\mathbf{\mathcal{G}}_{t}$, $\mathbf{\mathbf{\theta}}^{+}$
is the next-stage discrete state vector with its transition function
$\mathbf{\mathcal{H}}_{t}$, $\mathbf{\omega}$ and $\mathbf{\epsilon}$
are two vectors of random variables, and $U_{t}(\mathbf{x},\mathbf{a})$
is the utility function, $\beta$ is the discount factor, and $\mathbb{E}_{t}\{\cdot\}$
is the expectation operator conditional on information at time $t$.

The following is the algorithm of the numerical DP with value function
iteration for finite horizon problems.
\begin{description}
\item [{Algorithm}] 1. \textit{Numerical Dynamic Programming with Value
Function Iteration for Finite Horizon Problems} 
\item [{\textit{Initialization.}}] \textit{Choose the approximation nodes,
$\mathbb{X}=\{\mathbf{x}_{i}:$ $1\leq i\leq M\}\subset\mathbb{R}^{k}$,
for every $t<T$, and choose a functional form for $\hat{V}(\mathbf{x},\mathbf{\mathbf{\theta}};{\bf b})$.
Let $\hat{V}(\mathbf{x},\mathbf{\mathbf{\theta}};{\bf b}_{T})\equiv V_{T}(\mathbf{x},\mathbf{\mathbf{\theta}})$.
Then for $t=t_{N-1},t_{N-2},\ldots,t_{0}$, iterate through steps
1 and 2. } 
\item [{\textit{Step}}] \textit{1. Maximization Step. Compute 
\begin{eqnarray*}
 &  & v_{i,j}=\max_{\mathbf{a}\in{\cal D}(\mathbf{x}_{i},\mathbf{\mathbf{\theta}}_{j},t)}\ U_{t}(\mathbf{x}_{i},\mathbf{a})+\beta\mathbb{E}\left\{ \hat{V}(\mathbf{x}^{+},\mathbf{\mathbf{\theta}}^{+};{\bf b}_{t+\Delta t})\right\} \\
 &  & \text{{\rm \ \ \ \ \ \ \ \ \ \ \ \ \ s.t. \ \ \ }}\mathbf{x}^{+}=\mathbf{\mathcal{G}}_{t}(\mathbf{x}_{i},\mathbf{\mathbf{\theta}}_{j},\mathbf{a},\mathbf{\omega}),\\
 &  & \text{{\rm \ \ \ \ \ \ \ \ \ \ \ \ \ \ \ \ \ \ \ \ \ \ }}\mathbf{\mathbf{\theta}}^{+}=\mathbf{\mathcal{H}}_{t}(\mathbf{\mathbf{\theta}}_{j},\mathbf{\epsilon}),
\end{eqnarray*}
for each $\mathbf{\mathbf{\theta}}_{j}\in\Theta$, $\mathbf{x}_{i}\in\mathbb{X}$,
$1\leq i\leq M.$ } 
\item [{\textit{Step}}] \textit{2. Fitting Step. Using an appropriate approximation
method, compute the ${\bf b}_{t}$ such that $\hat{V}(\mathbf{x},\mathbf{\mathbf{\theta}}_{j};{\bf b}_{t})$
approximates $(\mathbf{x}_{i},v_{i,j})$ data for each $\mathbf{\mathbf{\theta}}_{j}\in\Theta$.
} 
\end{description}
Algorithm 1 shows that there are three main components in value function
iteration for DP problems: optimization, approximation, and integration. 

\subsection{Optimization}

We apply some fast Newton-type solvers, such as the NPSOL optimization
package, to solve the maximization problems in the numerical DP algorithm
(Algorithm 1). The maximization problem (\ref{Eq:Port_Power_Model})
is formulated in terms of $2k$ control variables ($\mathbf{\delta}_{t}^{+}$
and $\mathbf{\delta}_{t}^{-}$), and $2k$ bound constraints ($\mathbf{\delta}_{t}^{+},\mathbf{\delta}_{t}^{-}\geq0$),
$(k+1)$ inequality constraints for no-shorting and no-borrowing,
and other unknowns are expressed in terms of the controls, where $k$
is the number of risky assets. The problem (\ref{Eq:Port_Power_Cons_Model})
has one more control at each time, $c_{t}$.

Parallelization allows researchers to solve huge problems and is the
foundation of modern scientific computation. Our work shows that parallelization
can also be used effectively in solving the dynamic portfolio optimization
problems using the numerical DP method. The key fact is that at each
maximization step, there are many independent optimization problems,
one for each $(\mathbf{x}_{i},\mathbf{\theta}_{j})$. In our portfolio
problems there are often thousands or millions of such independent
problems. See Cai et al. (2015) for a more detailed discussion of
parallel DP methods.

\subsection{Numerical Approximation}

An approximation scheme consists of two parts: basis functions and
approximation nodes. Approximation nodes can be chosen as uniformly
spaced nodes, Chebyshev nodes, or some other specified nodes. From
the viewpoint of basis functions, approximation methods can be classified
as either spectral methods or finite element methods. A spectral method
uses globally nonzero basis functions $\phi_{j}(\mathbf{x})$ such
that $\hat{V}(\mathbf{x};{\bf b})=\sum_{j}b_{j}\phi_{j}(\mathbf{x})$.
Examples of spectral methods include ordinary polynomial approximation,
Chebyshev polynomial approximation, and shape-preserving Chebyshev
polynomial approximation (Cai and Judd, 2013), and Hermite approximation
(Cai and Judd, 2015). In contrast, a finite element method uses locally
nonzero basis functions $\phi_{j}(\mathbf{x})$, which are nonzero
over sub-domains of the approximation domain. Examples of finite element
methods include piecewise linear interpolation, cubic splines, B-splines,
and shape-preserving rational splines (Cai and Judd, 2012). See Cai
(2010, 2019) and Judd (1998) for more details.

We prefer Chebyshev polynomials when the value function is smooth.
Chebyshev basis polynomials on $[-1,1]$ are defined as $\mathcal{T}_{j}(x)=\cos(j\cos^{-1}(x)),$
while general Chebyshev basis polynomials on $[x_{\min},x_{\max}]$
are defined as $\mathcal{T}_{j}((2x-x_{\min}-x_{\max})/(x_{\max}-x_{\min}))$
for $j=0,1,2,\ldots$. For Chebyshev approximation, we know Chebyshev
nodes are the most efficient approximation nodes, and it is often
enough to let the number of Chebyshev nodes, $m$, be equal to the
number of unknown Chebyshev coefficients for one-dimensional problems;
that is, $m=d+1$, where $d$ is the degree of Chebyshev polynomials.
For one-dimensional problems using $m$ approximation nodes in a state
space $[x_{\min},x_{\max}]$, the corresponding Chebyshev nodes are
\[
x_{i}=(z_{i}+1)(x_{\max}-x_{\min})/2+x_{\min}
\]
with $z_{i}=-\cos\left((2i-1)\pi/(2m)\right)$ for $i=1,...,m$. If
we have Lagrange data $\{(x_{i},v_{i}):$ $i=1,\ldots,m\}$ with $v_{i}=V(x_{i})$,
then $V$ can be approximated by a degree-$d$ Chebyshev polynomial
\begin{equation}
\hat{V}(x;\mathbf{b})=\sum_{j=0}^{d}b_{j}\mathcal{T}_{j}\left(\frac{2x-x_{\min}-x_{\max}}{x_{\max}-x_{\min}}\right),\label{eq:ChebyApprox}
\end{equation}
where $b_{j}$ can be computed with the Chebyshev regression algorithm,
that is, 
\begin{equation}
b_{j}=\frac{2}{m}\sum_{i=1}^{m}v_{i}\mathcal{T}_{j}(z_{i}),\text{ \ \ \ \ }j=1,\ldots,d,\label{eq:ChebyCoefs-1}
\end{equation}
and $b_{0}=\sum_{i=1}^{m}v_{i}/m$.

In a $k$-dimensional approximation problem, let the domain of the
value function be 
\[
\left\{ \mathbf{x}=(x_{1},\ldots,x_{k}):\,x_{\min,j}\leq x_{j}\leq x_{\max,j},\,j=1,\ldots k\right\} ,
\]
for some real numbers $x_{\min,j}$ and $x_{\max,j}$, with $x_{\max,j}>x_{\min,j}$
for $j=1,\ldots,k$. Let $\mathbf{x}_{\min}=(x_{\min,1},\ldots,x_{\min,k})$
and $\mathbf{x}_{\max}=(x_{\max,1},\ldots,x_{\max,k})$. Then we denote
$[\mathbf{x}_{\min},\mathbf{x}_{\max}]$ as the hyper-rectangle domain.
Let $\mathbf{\alpha}=(\alpha_{1},\ldots,\alpha_{k})$ be a vector
of nonnegative integers. Let $\mathcal{T}_{\mathbf{\alpha}}(\mathbf{z})$
denote the product $\prod_{1\leq j\leq k}\mathcal{T}_{\alpha_{j}}(z_{j})$
for $\mathbf{z}=(z_{1},\ldots,z_{k})\in[-1,1]^{k}$. Let 
\[
\mathbf{Z}(\mathbf{x})=\left(\frac{2x_{1}-x_{\min,1}-x_{\max,1}}{x_{\max,1}-x_{\min,1}},\ldots,\frac{2x_{d}-x_{\min,k}-x_{\max,k}}{x_{\max,k}-x_{\min,k}}\right)
\]
for any $\mathbf{x}=(x_{1},\ldots,x_{k})\in[\mathbf{x}_{\min},\mathbf{x}_{\max}]$.

Using these notations, the degree-$d$ complete Chebyshev approximation
for $V(\mathbf{x})$ is 
\begin{equation}
\hat{V}_{d}(\mathbf{x};\mathbf{b})=\sum_{0\leq|\mathbf{\alpha}|\leq d}b_{\mathbf{\alpha}}\mathcal{T}_{\mathbf{\alpha}}\left(\mathbf{Z}(\mathbf{x})\right),\label{eq:compCheby}
\end{equation}
where $|\mathbf{\alpha}|\equiv\sum_{j=1}^{k}\alpha_{j}$ for the nonnegative
integer vector $\mathbf{\alpha}=(\alpha_{1},\ldots,\alpha_{k})$.
So the number of terms with $0\leq|\mathbf{\alpha}|\leq d$ is $\binom{d+k}{k}$
for the degree-$d$ complete Chebyshev approximation in $\mathbb{R}^{k}$.
Note that the complete Chebyshev approximation does not suffer from
the so-called ``curse of dimensionality''. For example, a degree-2
complete Chebyshev approximation in a 100-dimensional state space
has only 5,151 terms.

For $k$-dimensional problems in the state space $[\mathbf{x}_{\min},\mathbf{x}_{\max}]$,
if we use $m$ grids in each dimension, then there are $m^{k}$ approximation
nodes with the tensor grid, and the values of Chebyshev nodes in dimension
$j$ are
\[
x_{i,j}=(z_{i}+1)(x_{\max,j}-x_{\min,j})/2+x_{\min,j}.
\]
Usually we can just choose $m=d+1$, as a higher number of nodes has
little improvement on accuracy in approximation.

Using the Chebyshev nodes as approximation nodes $\mathbf{x}_{i}=(x_{i,1},...,x_{i,k})$
and their corresponding values $V(\mathbf{x}_{i})=v_{i}$, we can
compute Chebyshev coefficients directly using a multi-dimensional
Chebyshev regression algorithm. That is, 
\begin{equation}
b_{\mathbf{\alpha}}=\frac{2^{k-n}}{m^{k}}\sum_{i=1}^{m^{k}}v_{i}\mathcal{T}_{\alpha}(\mathbf{Z}(\mathbf{x}_{i})),\label{eq:ChebyCoefs-1-1}
\end{equation}
for any nonnegative integer vector $\mathbf{\alpha}=(\alpha_{1},\ldots,\alpha_{k})$
with $0\leq\alpha_{j}\leq d$, where $n$ is the number of zero elements
in $\mathbf{\alpha}$. 

\subsection{Numerical Integration}

In the objective function of the Bellman equations, we often need
to compute the conditional expectation of $V(\mathbf{x}^{+}\mid\mathbf{x},\mathbf{a})$.
When the random variable is continuous, we have to use numerical integration
to compute the expectation.

One naive way is to apply Monte Carlo or pseudo Monte Carlo methods
to compute the expectation. By the Central Limit Theorem, the numerical
error of the integration computed by (pseudo) Monte Carlo methods
has a distribution that is close to normal, and so no bound for this
numerical error exists. Moreover, the optimization problem often needs
hundreds or thousands of evaluations of the objective function. This
implies that once one evaluation of the objective function has a big
numerical error, the previous iterations to solve the optimization
problem may make no sense, and the iterations may never converge to
the optimal solution. Thus it is not practical to apply (pseudo) Monte
Carlo methods to the optimization problem generally, unless the stopping
criterion of the optimization problem is set very loosely.

Therefore, it will be better to have a numerical integration method
with a bounded numerical error. Here we present a common numerical
integration method to use when the random variable is normal.

In the expectation operator of the objective function of the Bellman
equation, if the random variable has a normal distribution, then it
will be good to apply the Gauss-Hermite quadrature formula to compute
the numerical integration. That is, if we want to compute $\mathbb{E}\{f(Y)\}$
where $Y$ has a distribution $\mathcal{N}(\mu,\sigma^{2})$, then
\begin{eqnarray*}
\mathbb{E}\{f(Y)\} & = & (2\pi\sigma^{2})^{-1/2}\int_{-\infty}^{\infty}f(y)e^{-(y-\mu)^{2}/(2\sigma^{2})}dy\\
 & = & (2\pi\sigma^{2})^{-1/2}\int_{-\infty}^{\infty}f(\sqrt{2}\,\sigma\,x+\mu)e^{-x^{2}}\sqrt{2}\sigma dx\\
 & \doteq & \pi^{-{\frac{1}{2}}}\sum_{i=1}^{m}\omega_{i}f(\sqrt{2}\sigma x_{i}+\mu),
\end{eqnarray*}
where $\omega_{i}$ and $x_{i}$ are the Gauss-Hermite quadrature
weights and nodes over $(-\infty,\infty)$, and $m$ is the number
of quadrature nodes.

If $Y$ is log normal, i.e., $\log(Y)$ has a distribution $\mathcal{N}(\mu,\sigma^{2})$,
then we can assume that $Y=e^{X}$ where $X\sim\mathcal{N}(\mu,\sigma^{2})$,
thus 
\[
\mathbb{E}\{f(Y)\}=\mathbb{E}\{f(e^{X})\}\doteq\pi^{-{\frac{1}{2}}}\sum_{i=1}^{m}\omega_{i}f\left(e^{\sqrt{2}\sigma x_{i}+\mu}\right).
\]

If we want to compute a multidimensional integration, we could apply
the product rule. For example, suppose that we want to compute $\mathbb{E}\{f(\mathbf{X})\}$,
where $\mathbf{X}$ is a random vector with multivariate normal distribution
$\mathcal{N}(\text{\ensuremath{\mu}},\Sigma)$ over $(-\infty,+\infty)^{k}$,
where $\mathbf{\mu}$ is the mean column vector and $\Sigma$ is the
covariance matrix, then we could do the Cholesky factorization first,
i.e., find a lower triangular matrix $\mathbf{L}$ such that $\Sigma=\mathbf{L}\mathbf{L}^{\top}$.
This is feasible as $\Sigma$ must be a positive semi-definite matrix,
from the covariance property. Thus, 
\begin{eqnarray*}
\mathbb{E}\{f(\mathbf{X})\} & = & \left((2\pi)^{k}{\det}(\Sigma)\right)^{-1/2}\int_{R^{k}}f(\mathbf{y})e^{-(\mathbf{y}-\mathbf{\mu})^{\top}\Sigma^{-1}(\mathbf{y}-\mathbf{\mu})/2}d\mathbf{y}\\
 & = & \left((2\pi)^{d}{\det}(\mathbf{L})^{2}\right)^{-1/2}\int_{R^{k}}f\left(\sqrt{2}\mathbf{L}\mathbf{x}+\mathbf{\mu}\right)e^{-\mathbf{x}^{\top}\mathbf{x}}2^{k/2}{\det}(\mathbf{L})d\mathbf{x}\\
 & \doteq & \pi^{-{\frac{k}{2}}}\sum_{i_{1}=1}^{m}\cdots\sum_{i_{k}=1}^{m}\omega_{i_{1}}\cdots\omega_{i_{k}}f\Bigg(\sqrt{2}L_{1,1}x_{i_{1}}+\mu_{1},\\
 &  & \sqrt{2}(L_{2,1}x_{i_{1}}+L_{2,2}x_{i_{2}})+\mu_{2},\cdots,\sqrt{2}(\sum_{j=1}^{k}L_{k,j}x_{i_{j}})+\mu_{k}\Bigg),
\end{eqnarray*}
where $\omega_{i}$ and $x_{i}$ are the Gauss-Hermite quadrature
weights and nodes over $(-\infty,\infty)$, $L_{i,j}$ is the $(i,j)$-element
of $\mathbf{L}$, and $\det(\cdot)$ means the matrix determinant
operator.

\section{Problems with Stochastic Parameters\label{sec:Problems-with-Stochastic}}

In this section, we solve the three-asset with-consumption model (\ref{Eq:Port_Power_Cons_Model})
with either a stochastic interest rate or volatility. We use weekly
time periods (i.e., $\Delta t=1/52$ years) in $T=3$ years (so the
number of periods is $N=156$), and the discount rate is 0.05. In
the examples, the assets available for trading include one risk-free
asset with an interest rate $r$ and $k=2$ uncorrelated risky assets
with log-normal returns. We assume that the utility function is $U(C)=C^{1-\gamma}/(1-\gamma)$
with $\gamma=3$. The terminal value function is 
\[
V_{T}(W,\mathbf{x},\text{\ensuremath{\theta}})=\frac{(r(1-\tau\mathbf{e}^{\top}\mathbf{x})W)^{1-\gamma}\Delta t}{(1-\gamma)(1-\beta)}=W^{1-\gamma}\cdot G_{T}(\mathbf{x},\mathbf{\theta})
\]
where 
\[
G_{T}(\mathbf{x},\mathbf{\theta}):=\frac{(r(1-\tau\mathbf{e}^{\top}\mathbf{x}))^{1-\gamma}\Delta t}{(1-\gamma)(1-\beta)}.
\]
The proportional transaction cost ratio is $\tau=0.1\%$ for buying
or selling risky assets. The default parameter values are $r=0.03$,
$\mu=(0.07,0.07)^{\top}$, and $\sigma=(0.2,0.2)^{\top}$, if they
are not stochastic. In the numerical DP method, we choose degree-$60$
complete Chebyshev polynomials to approximate value functions, use
$61^{2}$ tensor Chebyshev nodes as the approximation nodes, and implement
the multi-dimensional product Gauss-Hermite quadrature rule with $3^{2}$
tensor quadrature nodes.

\subsection{Stochastic Interest Rate}

We solve a case with a stochastic interest rate $r_{t}$. The interest
rate $r_{t}$ is assumed to be a Markov chain. It has three possible
values: $r^{1}=0.03$, $r^{2}=0.04$, $r^{3}=0.05$. That is, $r_{t}$
is the discrete state $\theta_{t}$ in the model (\ref{Eq:Port_Power_Cons_Model}).
Its transition probability matrix is 
\[
\left[\begin{array}{ccc}
0.6 & 0.4 & 0\\
0.2 & 0.6 & 0.2\\
0 & 0.4 & 0.6
\end{array}\right],
\]
where its $(i,j)$ element represents the transition probability from
$r_{t}=r^{i}$ to $r_{t+\Delta t}=r^{j}$, for $i,j=1,2,3$.

\begin{figure}
\begin{centering}
\includegraphics[width=0.5\textwidth]{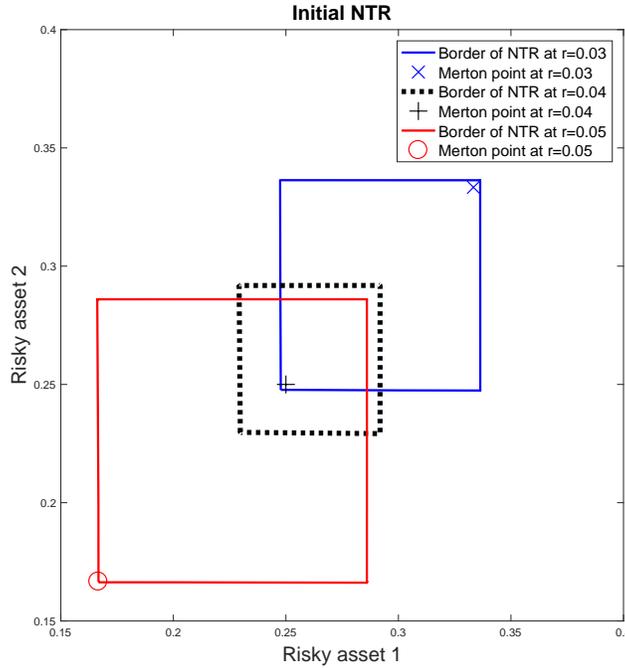}
\par\end{centering}
\caption{\label{fig:StoPara-r}NTRs with a stochastic interest rate.}
\end{figure}

Figure \ref{fig:StoPara-r} shows the NTRs at the initial time with
different initial interest rates. In the figure, the mark, the plus,
and the circle represent the Merton points at three discrete state
values: $r=0.03$, $0.04$, and $0.05$, respectively, if we assume
that the interest rate is fixed at its initial value. The top-right
square, the middle square, and the left-bottom square represents the
NTRs at the three discrete state values: $r=0.03$, $0.04$, and $0.05$,
respectively. We see that a higher interest rate implies a NTR and
a Merton point closer to the origin of the coordinate system. This
is consistent with the formula of the Merton point, (\ref{eq:Merton_point}).
Moreover, the Merton point at $r=0.03$ or $0.05$ stays away from
the center of its corresponding square, and the ``outside edges''
of its NTR are closer to the point than the ``inside edges''. That
is, the circle is located in the left-bottom corner of its corresponding
NTR, and the mark is located in the top-right corner of its corresponding
NTR. If the discrete state at time $t$ is $r=0.05$, then a portfolio
before rebalancing that is located near the center of all the three
squares (i.e., a point in the northeast direction of the circle in
the bottom-left square) tends to have less trade than another portfolio
that is the same distance from the circle but further away from the
center (i.e., a point in the southwest direction of the circle in
the bottom-left square). This confirms the intuition that if the portfolio
is in a position closer to the center of all NTRs, there is little
or no incentive to trade, because the expected direction of next period's
trade is close to zero.

\subsection{Stochastic Volatility}

Now we examine the case with stochastic volatility of risky returns,
$\sigma_{t}=(\sigma_{t,1},\sigma_{t,2})^{\top}$. We assume that $\sigma_{t,1}$
and $\sigma_{t,2}$ are discrete Markov chains and independent of
each other. Each $\sigma_{t,i}$ has two possible values: $0.16$
and 0.24, and its transition probability matrix is 
\[
\left[\begin{array}{cc}
0.75 & 0.25\\
0.25 & 0.75
\end{array}\right]
\]
for each $i=1,2$.

\begin{figure}
\begin{centering}
\includegraphics[width=0.5\textwidth]{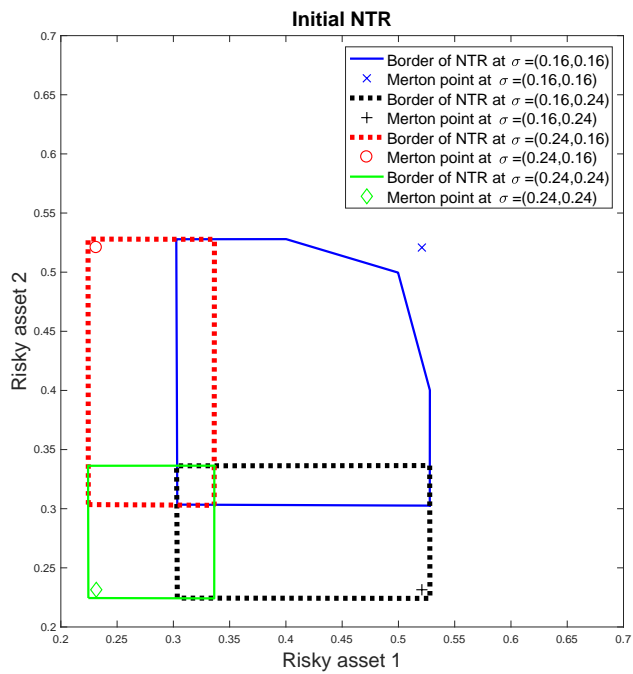}
\par\end{centering}
\caption{\label{fig:StoPara_sig}NTRs with stochastic $\sigma$.}
\end{figure}

Figure \ref{fig:StoPara_sig} displays the NTRs for four possible
discrete states of $(\sigma_{t,1},\sigma_{t,2})$ at the initial time.
The top-right polygon is the NTR for the state $(\sigma_{t,1},\sigma_{t,2})=(0.16,0.16)$,
the bottom-left square is the NTR for the state $(\sigma_{t,1},\sigma_{t,2})=(0.24,0.24)$,
and the top-left and the bottom-right squares are respectively the
NTRs for the states $(\sigma_{t,1},\sigma_{t,2})=(0.24,0.16)$ and
$(\sigma_{t,1},\sigma_{t,2})=(0.16,0.24)$. The mark, the diamond,
the circle, and the plus are the corresponding Merton points if we
assume that $(\sigma_{t,1},\sigma_{t,2})$ are fixed at their initial
values. We see that a higher $\sigma_{t,i}$ implies a NTR and a Merton
point closer to the origin of the coordinate system, consistent with
the formula of the Merton point, (\ref{eq:Merton_point}). Moreover,
when $(\sigma_{t,1},\sigma_{t,2})=(0.24,0.24)$, the no-shorting and/or
no-borrowing constraints become binding for some portfolios before
rebalancing. For example, if the portfolio before re-allocation is
(0.5,0.5), the top-right vertex of the top-right polygon, then the
optimal trading strategy is to keep the current portfolio unchanged
(i.e., the no-borrowing constraint is binding).

\end{document}